%% file: main.tex
\documentclass[camera]{jpaper}
\usepackage{makecell}

\usepackage[nocompress]{cite}
\usepackage{soul}

\usepackage{amsmath}
\usepackage{graphicx}
\usepackage[linesnumbered,ruled]{algorithm2e}

\usepackage{algpseudocode}
\usepackage{titlesec}
\usepackage{anyfontsize}

\usepackage{datetime}
\algrenewcommand\algorithmiccomment[2][\normalsize]{{#1\hfill\(\triangleright\) #2}}
\usepackage{geometry} \titlespacing*{\section}{0pt}{2pt plus 0.5pt minus 0.5pt}{0pt}
\titlespacing*{\subsection}{0pt}{2pt plus 0.5pt minus 0.5pt}{0pt}
\titlespacing*{\subsubsection}{0pt}{2pt plus 0.5pt minus 0.5pt}{0pt}

\makeatletter
\renewcommand{\@makefnmark}{\hbox{\textsuperscript{\scriptsize{\@thefnmark}}}}
\makeatother

\usepackage{array}
\newcolumntype{L}[1]{>{\raggedright\let\newline\\\arraybackslash\hspace{0pt}}m{#1}}
\newcolumntype{C}[1]{>{\centering\let\newline\\\arraybackslash\hspace{0pt}}m{#1}}
\newcolumntype{R}[1]{>{\raggedleft\let\newline\\\arraybackslash\hspace{0pt}}m{#1}}
\usepackage{bm}

\makeatletter
\let\MYcaption\@makecaption
\makeatother

\usepackage[font=footnotesize]{subcaption}

\makeatletter
\let\@makecaption\MYcaption
\makeatother

\usepackage{fixltx2e}
\usepackage{dblfloatfix}
\usepackage[nolessnomore, italic]{mathastext}
\usepackage[T1]{fontenc}
\usepackage[usenames,dvipsnames,svgnames,table]{xcolor}
\usepackage{multirow}
\usepackage{hhline}
\usepackage[normalem]{ulem}
\usepackage{setspace}
\usepackage{indentfirst}
\usepackage{footmisc}
\usepackage{pifont}% http://ctan.org/pkg/pifont

\newcommand{\affilETH}[0]{\textsuperscript{$\dagger$}}
\newcommand{\affilCMU}[0]{\textsuperscript{$\ddagger$}}
\newcommand{\affilUIUC}[0]{\textsuperscript{$\diamond$}}
\newcommand{\affilNUDT}[0]{\textsuperscript{$\star$}}
\newcommand{\affilIPM}[0]{\textsuperscript{$\S$}}
\newcommand{\affilCUHK}[0]{\textsuperscript{$\odot$}}

\definecolor{amber}{rgb}{1.0, 0.49, 0.0}
\definecolor{darkgreen}{rgb}{0.0, 0.2, 0.13}
\definecolor{darkbyzantium}{rgb}{0.36, 0.22, 0.33}
\definecolor{darkseagreen}{rgb}{0.56, 0.74, 0.56}
\definecolor{darkspringgreen}{rgb}{0.09, 0.45, 0.27}
\definecolor{dollarbill}{rgb}{0.52, 0.73, 0.4}

\newif\ifcameraready
\camerareadytrue
\newcommand{\versionnum}[0]{6.4} 

\newcommand*\circled[1]{\tikz[baseline=(char.base)]{
            \node[shape=circle,draw,inner sep=0pt,fill=black, text=white] (char) {#1};}}
\newcommand{\substrate}[1]{FIGARO}
\newcommand{\cachename}[1]{FIGCache}
\newcommand{\relcommand}[1]{\texttt{RELOC}}

\newcommand{\chyhw}[0]{}
\newcommand{\chyhwI}[0]{}
\newcommand{\chyhwII}[0]{}
\newcommand{\sg}[0]{}
\newcommand{\sgx}[0]{}
\newcommand{\sgi}[0]{}
\newcommand{\sgii}[0]{}
\newcommand{\sgiii}[0]{}
\ifcameraready
  \newcommand{\sgiv}[0]{}
  \newcommand{\sgv}[0]{}
\else
  \newcommand{\sgiv}[1]{\textcolor{MidnightBlue}{#1}}
  \newcommand{\sgv}[1]{\textcolor{BrickRed}{#1}}
\fi

\begin{document}
\bstctlcite{IEEEexample:BSTcontrol} % for controls on # authors and so forth

%\title{\mechanism: Violating DRAM Timing Constraints \\ for High-Throughput True Random Number Generation \\ using Commodity DRAM Devices} 

\title{FIGARO: Improving System Performance \\ via Fine-Grained In-DRAM Data Relocation and Caching} %should we change FIGARO to FIGAMO? Migration also contain 'R' though.

%\title{\mechanism: Using Commodity DRAM Devices \\
%to Generate True Random Numbers \\ at Low Latency and High Throughput} 

%
% paper title
% can use linebreaks \\ within to get better formatting as desired

% author names and affiliations
% use a multiple column layout for up to two different
% affiliations

\author{\fontsize{11.25}{13}\selectfont
{Yaohua Wang\affilNUDT}\quad%
{Lois Orosa\affilETH}\quad%
{Xiangjun Peng\affilCUHK\affilNUDT}\quad%
{Yang Guo\affilNUDT}\quad%
{Saugata Ghose\affilUIUC\affilCMU}\quad%
{Minesh Patel\affilETH}\\%
{\fontsize{11.25}{13}\selectfont%
{Jeremie S. Kim\affilETH}\quad%
{Juan Gómez Luna\affilETH}\quad%
{Mohammad Sadrosadati\affilIPM}\quad%
{Nika Mansouri Ghiasi\affilETH}\quad%
{Onur Mutlu\affilETH\affilCMU}\vspace{-8pt}}\\\\% 
{\fontsize{9.9}{12}\selectfont\emph{%
\affilNUDT National University of Defense Technology\quad%
\affilETH ETH Z{\"u}rich\quad%
\affilCUHK Chinese University of Hong Kong}}\\%
{\fontsize{9.9}{12}\selectfont{\emph{%
\affilUIUC University of Illinois at Urbana--Champaign~~~%
\affilCMU Carnegie Mellon University~~~%
\affilIPM Institute of Research in Fundamental Sciences}}%
}%
}

% conference papers do not typically use \thanks and this command
% is locked out in conference mode. If really needed, such as for
% the acknowledgment of grants, issue a \IEEEoverridecommandlockouts
% after \documentclass

% for over three affiliations, or if they all won't fit within the width
% of the page, use this alternative format:
% 
%\author{\IEEEauthorblockN{Michael Shell\IEEEauthorrefmark{1},
%Homer Simpson\IEEEauthorrefmark{2},
%James Kirk\IEEEauthorrefmark{3}, 
%Montgomery Scott\IEEEauthorrefmark{3} and
%Eldon Tyrell\IEEEauthorrefmark{4}}
%\IEEEauthorblockA{\IEEEauthorrefmark{1}School of Electrical and Computer Engineering\\
%Georgia Institute of Technology,
%Atlanta, Georgia 30332--0250\\ Email: see http://www.michaelshell.org/contact.html}
%\IEEEauthorblockA{\IEEEauthorrefmark{2}Twentieth Century Fox, Springfield, USA\\
%Email: homer@thesimpsons.com}
%\IEEEauthorblockA{\IEEEauthorrefmark{3}Starfleet Academy, San Francisco, California 96678-2391\\
%Telephone: (800) 555--1212, Fax: (888) 555--1212}
%\IEEEauthorblockA{\IEEEauthorrefmark{4}Tyrell Inc., 123 Replicant Street, Los Angeles, California 90210--4321}}

% use for special paper notices
%\IEEEspecialpapernotice{(Invited Paper)}

% make the title area
\maketitle
\thispagestyle{plain} 
\pagestyle{plain}

\setstretch{0.86}
%\setstretch{1}
\renewcommand{\footnotelayout}{\setstretch{0.9}}

% SAUGATA: add page numbers
%\ifcameraready
%\else
%  \thispagestyle{plain}
%  \pagestyle{plain}
%\fi

% Metadata Information
%\newcommand{\versionnum}[0]{2.29 12/27/2020 6:15PM EST}

\fancyhead{}
\ifcameraready
 \thispagestyle{plain}
 \pagestyle{plain}
 % \pagenumbering{gobble}
\else
 \fancyhead[C]{\textcolor{MidnightBlue}{\emph{Version \versionnum~---~\today, \ampmtime}}}
 \fancypagestyle{firststyle}
 {
   \fancyhead[C]{\textcolor{MidnightBlue}{\emph{Version \versionnum~---~\today, \ampmtime}}}
   \fancyfoot[C]{\thepage}
 }
 \thispagestyle{firststyle}
 \pagestyle{firststyle}
\fi

%\setstretch{0.80}
%\renewcommand{\footnotelayout}{\setstretch{0.9}}

%\input{sections/abstract}

% \begin{IEEEkeywords}
% \todo[do we need them?]
% \end{IEEEkeywords}

% For peer review papers, you can put extra information on the cover
% page as needed:
% \ifCLASSOPTIONpeerreview
% \begin{center} \bfseries EDICS Category: 3-BBND \end{center}
% \fi
%
% For peerreview papers, this IEEEtran command inserts a page break and
% creates the second title. It will be ignored for other modes.
% \IEEEpeerreviewmaketitle

%%%%%% -- PAPER CONTENT STARTS-- %%%%%%%%

\input{0_abstract}

\input{1_introduction}
\input{2_background}
\input{3_existing_in_DRAM_Cache}
\input{4_Relo_Cab_Design}
\input{5_FG_Cache_Design}
\input{6_Methodology}
\input{7_evaluation}

\input{8_Sensitivity}
\input{9_related}

\input{10_conclusion}

%%%%%%% -- PAPER CONTENT ENDS -- %%%%%%%%

% use section* for acknowledgement
%\section*{Acknowledgments} \jksix{We thank Ivan Puddu for useful comments and
%anonymous reviewers and SAFARI group members for feedback.} 
%\section*{Acknowledgments}
%We thank the anonymous reviewers for their feedback.
%This work is partially supported by the Intel Science and
%Technology Center, the CMU Data Storage Systems Center,
%NSF grants 1212962/1320531, and gifts from Intel and
%Seagate.
% \todo[check this]
\section*{Acknowledgments} 
We thank the anonymous reviewers, SAFARI group members for the feedback and the stimulating research environment. This work was supported by \sgi{a Hunan Province Science and Technology Planning project} (No.\ 2019RS2027), \sgi{a National University of Defense Technology research project} \chyhwI{(No.\ 18/19-QNCXJ-WYH),}
%\todo{what is the official name of this?}, 
\chyhw{and the industrial partners of SAFARI, \sgi{especially Google,} Huawei, Intel, Microsoft, and VMware.}

% trigger a \newpage just before the given reference
% number - used to balance the columns on the last page
% adjust value as needed - may need to be readjusted if
% the document is modified later
%\IEEEtriggeratref{8}
% The "triggered" command can be changed if desired:
%\IEEEtriggercmd{\enlargethispage{-5in}}

% references section

% can use a bibliography generated by BibTeX as a .bbl file
% BibTeX documentation can be easily obtained at:
% http://www.ctan.org/tex-archive/biblio/bibtex/contrib/doc/
% The IEEEtran BibTeX style support page is at:
% http://www.michaelshell.org/tex/ieeetran/bibtex/
%\bibliographystyle{IEEEtran}
% argument is your BibTeX string definitions and bibliography database(s)
%\bibliography{IEEEabrv,../bib/paper}
%
% <OR> manually copy in the resultant .bbl file
% set second argument of \begin to the number of references
% (used to reserve space for the reference number labels box)

%\balancecolumns
% That's all folks!

% \SetTracking
%  [ no ligatures = {f},
%  % spacing = {600*,-100*, },
%  % outer spacing = {450,250,150},
%  outer kerning = {*,*} ]
%  { encoding = * }
%  { -40 } % THIS CONTROLS HOW TIGHTLY TO SQUEEZE

\patchcmd{\thebibliography}{\clubpenalty4000}{\clubpenalty10000}{}{}     % no orphans
\patchcmd{\thebibliography}{\widowpenalty4000}{\widowpenalty10000}{}{}   % no widows
\patchcmd{\bibsetup}{\interlinepenalty=5000}{\interlinepenalty=10000}{}{} % no break of entry

{
%  \lsstyle % ENABLE ME IF YOU WISH TO SEE YOUR FIRSTBORN KILLED BEFORE YOUR VERY EYES 

  \setstretch{0.933}
  \let\OLDthebibliography\thebibliography
  \renewcommand\thebibliography[1]{
    \OLDthebibliography{#1}
    \setlength{\parskip}{0pt}
    \setlength{\itemsep}{0pt}
  }
  \bibliographystyle{IEEEtranS}
  \bibliography{ref}
}

%\input{appendix} 

% that's all folks
\end{document}

%% file: 0_abstract.tex
\begin{abstract} 
Main memory, \chyhw{composed} of DRAM, \chyhw{is} a performance bottleneck for many applications,
due to the high DRAM access latency.
In-DRAM caches work to mitigate this latency by augmenting regular-latency DRAM with
small-but-fast regions of DRAM that serve as a cache \chyhwI{for} the data
held in the regular-latency \chyhw{(i.e., slow)} region of DRAM.
While an effective \chyhwI{in-DRAM} cache can allow a large fraction of memory requests to be served
from a fast DRAM region, the latency savings are often hindered by
inefficient mechanisms for \sg{migrating (i.e., relocating)} copies of data into and out of the
\sg{fast} regions.
Existing in-DRAM caches have two sources of inefficiency:
(1)~their \chyhw{data relocation} granularity is an entire multi-kilobyte row of DRAM, 
even though much of the row may never be accessed due to poor data locality; and
(2)~because the relocation latency increases with the physical distance between
the \chyhw{slow} and fast regions, multiple fast regions are physically interleaved
among \chyhw{slow} regions to reduce the \chyhw{relocation} latency, resulting in increased
\chyhwI{hardware area and} manufacturing complexity.

% To improve DRAM latency, in-DRAM cache has been proposed by introducing heterogeneity into DRAM banks, where one region has a normal access latency and capacity, while the other has a fast access latency but small capacity, serving as an inclusive in-DRAM cache. However, such design suffers from the inefficient data relocation operations between fast and normal regions, where 1) the relocation granularity is an entire DRAM row, leading to limited performance improvement due to low data locality, and 2) the relocation latency increases with the increase of the physical distance between fast and normal regions, so that multiple fast regions are required and interleaved among normal regions to reduce data relocation latency, resulting in increased overhead and manufacturing complexity.  
	
We propose a new substrate, \substrate{}, that uses existing shared \chyhwI{global} buffers
\chyhwI{among subarrays} within a DRAM bank to provide \chyhwII{support for} \chyhw{in-DRAM} data relocation \chyhwI{across subarrays} at the granularity of a single
cache block.
\substrate{} has a distance-independent latency within a DRAM \chyhwI{bank}, and avoids complex
modifications to DRAM (such as the interleaving of fast and \chyhw{slow} regions).
%that can harm manufacturing yield.
Using \substrate{}, we design a fine-grained in-DRAM cache called
\chyhw{\cachename{}}.
The key idea of \chyhw{\cachename{}} is to cache only small, frequently-accessed 
portions of \chyhw{different} DRAM \chyhw{rows} in a designated region of DRAM.
By caching only the parts of \chyhw{each} row that are expected to be accessed
in the near future, we can pack more of the frequently-accessed data
into \cachename{}, and can benefit from additional row hits in DRAM
(i.e., accesses to an already-open row, which have a lower latency
than accesses to an unopened row).
\chyhw{\cachename{}} provides benefits for \chyhwI{systems with} both heterogeneous DRAM \chyhwI{banks}
(i.e., \chyhwI{banks} with fast regions and \chyhw{slow} regions) \emph{and}
conventional homogeneous DRAM \chyhwI{banks} (i.e., \chyhwI{banks} with only \chyhw{slow} regions).

% In this paper, we observe that the global row buffer in a DRAM bank is interconnected with all of the bank’s local row buffers. This enables us to build \substrate{}, which can perform cache-block level data relocation within a DRAM bank at a distance independent latency. Based on \substrate{}, we propose a fine-grained (i.e., sub-DRAM-row) in-DRAM cache (FG-Cache). The corresponding benefits are three-fold: 1) increased performance improvement, as multiple hot sub-DRAM-rows can be combined into one single DRAM row of the in-DRAM cache, the overall performance is substantially increased due to a higher in-DRAM cache hit and row buffer hit rate; 2) simplified in-DRAM cache design, fewer fast regions are needed and no longer necessary to be interleaved among normal regions; 3) potential benefit for conventional homogeneous DRAM chip, considerable performance gains can be achieved even by simply reserving a few DRAM rows in conventional DRAM chips to collocate hot sub-DRAM-rows.

Our evaluations across a wide variety of applications show that \chyhw{\cachename{}} improves the average performance of a system using DDR4 DRAM by 16.3\% and reduces average DRAM energy consumption by 7.8\% for 8-core workloads, over a conventional system without in-DRAM caching. We show that \chyhw{\cachename{}} outperforms state-of-the-art in-DRAM caching \chyhw{techniques}, and that its performance gains are robust across many system and mechanism parameters.
% We also discuss how FG-Cache can be used to prevent security attacks.
\end{abstract}

%% file: 1_introduction.tex
\section{Introduction}

\begin{figure}[h!] \centering
    \includegraphics[width=1\linewidth]{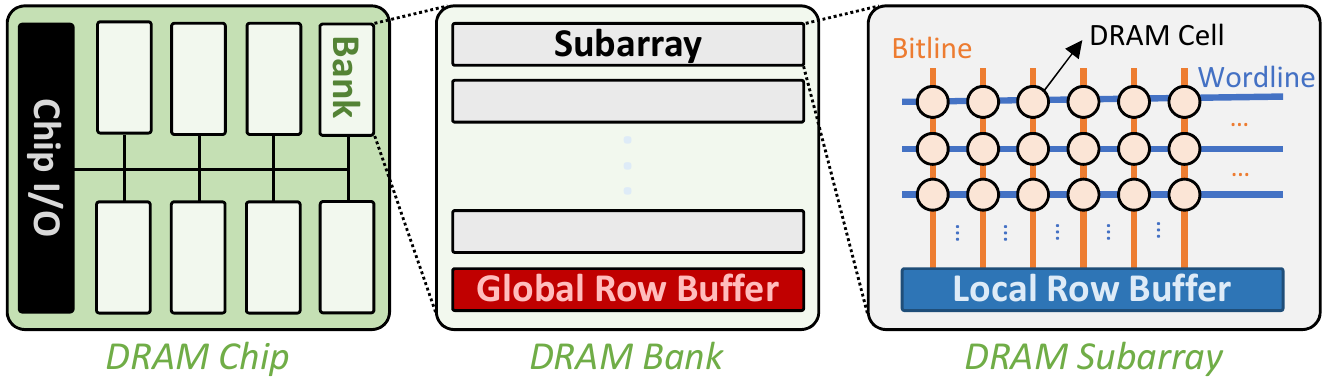}%
    \vspace{-0mm}
    \caption{\sgiii{Logical organization of a DRAM chip.}}
    \vspace{-0pt}
    \label{fig:organization}
\end{figure}

DRAM has long been the dominant technology for main memory systems. 
As many modern applications require greater amounts of DRAM to hold increasing amounts of data,
manufacturers are increasing the capacity of DRAM chips \chyhwI{via} manufacturing process technology scaling.
However, unlike capacity, DRAM access latency has not decreased significantly for decades, as latency improvements are traded off to instead decrease the cost-per-bit of DRAM\sgiii{~\cite{TieredDRAM,AsymBank,FLY-DRAM,AL-DRAM, mutlu.imw2013, RProbs}}.
To understand why, we study the high-level organization of a DRAM chip, as shown in Figure~\ref{fig:organization}.
The chip consists of multiple DRAM \emph{banks} (eight in DDR4 DRAM~\cite{JEDEC4}), where each bank is comprised of multiple homogeneous \emph{subarrays} (i.e., two-dimensional tiles)~\cite{SALP} of DRAM cells. 
Each DRAM cell stores a bit of data in the form of charge.
\sg{Reads and writes cannot be performed directly on the cell, as \chyhwII{the cell} holds only a limited amount of charge (in order to keep the cell area small), and this amount is too small to \chyhwII{drive the} I/O circuitry.
\sgii{Instead}, a cell in a subarray is connected via \chyhwII{a \emph{bitline}} to the subarray's \emph{local row buffer} (consisting of sense amplifiers)~\cite{SALP,DDMA}.  A local row buffer is used to sense, amplify, and hold the contents of one row of DRAM.
Each subarray has its own local row buffer, which consumes a relatively large area compared to a row of DRAM cells. To amortize this area and achieve low cost-per-bit, a commodity DRAM connects many DRAM cells to each sense amplifier on a single bitline (e.g., 512--2048 cells per bitline). \chyhwII{Doing so results in a long bitline} to accommodate the number of attached DRAM cells, and \chyhwII{a long bitline has} high parasitic capacitance. \chyhwII{Bitline} capacitance has a direct impact on DRAM access latency: the longer the bitline, the higher the parasitic capacitance, and, thus, the longer the latency required to bring the data from a row of DRAM cells into the local row buffer\chyhwII{~\cite{TieredDRAM}}.}
% As a DRAM cell can hold only a limited amount of charge (in order to keep the cell area small),
% a \emph{row buffer} \chyhw{(consisting of sense amplifiers)} is used to \chyhw{sense, amplify, and hold} the contents of one row of DRAM.
% Each subarray contains its own local row buffer~\cite{SALP,DDMA}.
% % , and all of the 
\sgii{The local row buffers in a bank are connected to a shared \emph{global row buffer}, which interfaces with the chip’s I/O drivers.}
% To mitigate the high area overhead of local row buffers and achieve low cost-per-bit, \chyhw{a commodity DRAM design connects} many DRAM cells to the local row buffer through long bitlines \chyhw{e.g., 512--2048} in each subarray. These bitlines have a high parasitic capacitance due to their long length, leading to high DRAM access latency. \chyhw{the longer the bitline length, the longer the access latency to bring the data from a row into the local row buffer.}
% % While accessing data that is already in the local row buffer (i.e., row buffer hit) is not affected by long bitlines, the row buffer hit rate is limited, and getting worse nowadays due to the interference among multi-cores~\cite{PartialActivation,SubrowBuffer,Micro-Pages}. 

\begin{figure}[h] \centering
\includegraphics[width=1.0\linewidth]{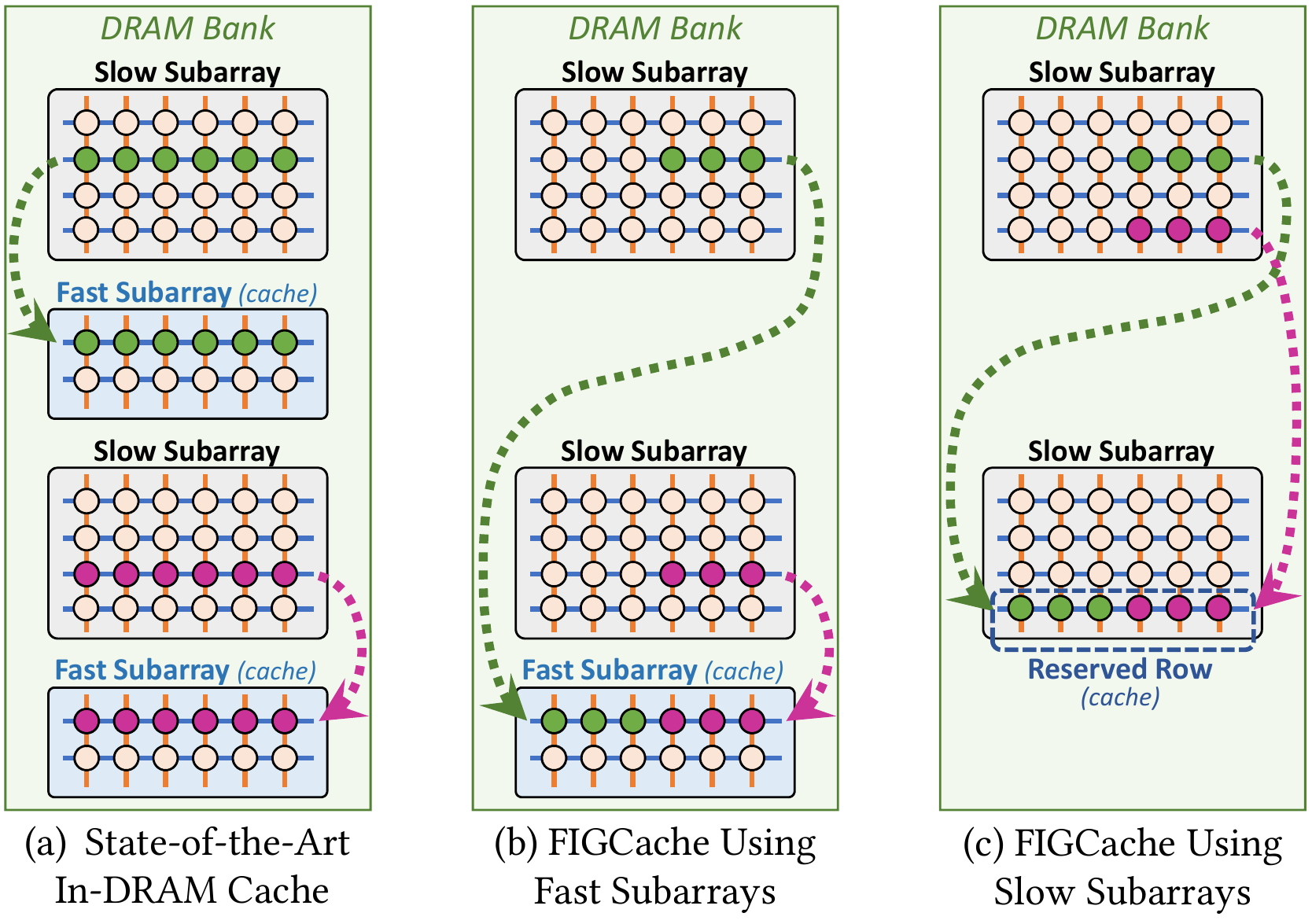}%
\vspace{-2pt}
\caption{\sgiii{(a)~State-of-the-art in-DRAM cache in a \sgiv{heterogeneous} bank with many fast subarrays interleaved among \chyhw{slow} subarrays;
(b)~\cachename{} in a \sgiv{heterogeneous} bank with fewer fast subarrays;
(c)~\cachename{} \sgiv{in} a conventional bank with no fast subarrays.}}
% \caption{\textbf{(a)} Due to the inefficient data relocation support, existing in-DRAM cache designs~\cite{DyAsy,LISA} 1) unnecessarily cache an entire DRAM row at a time, achieving limited performance gain due to low locality, and 2) employ many fast subarrays and interleave them among \chyhw{slow} subarrays to reduce data relocation latency, leading to high overhead and manufacturing complexity; \textbf{(b)} we propose fine-grained in-DRAM cache (FG-Cache) enabled by our efficient cacheblock relocation substrate (\substrate{}), FG-Cache reduces the caching granularity to a sub-DRAM-row, and employs fewer fast subarrays that are no longer necessary to be interleaved among \chyhw{slow} subarrays; \textbf{(c)} we can also build the in-DRAM cache by simply reserving a few DRAM rows in \chyhw{slow} subarrays, without introducing fast subarrays into memory banks, considerable performance gain can still be achieved by collocating multiple sub-DRAM-rows via \substrate{}.}
%\caption{In-DRAM Cache Designs}
\label{fig:in-DRAM-cache}
\vspace{-3pt}
\end{figure}

To improve DRAM latency while maintaining low cost-per-bit, prior works modify the DRAM organization to implement an \emph{in-DRAM cache}~\cite{DyAsy,LISA,AsymBank,TieredDRAM}.
\chyhwII{Many of these} works take the approach shown in Figure~\ref{fig:in-DRAM-cache}a, where they introduce \emph{heterogeneous subarrays} into DRAM banks.
In such a bank, one type of subarray (labeled a \emph{\chyhw{slow} subarray} in the figure) has a \chyhw{regular (i.e., slow)} access latency and \chyhw{large} capacity, while a second type of subarray (labeled \emph{fast subarray}) has a \chyhwI{low} access latency but small capacity (i.e., the subarray's bitlines are kept short to reduce parasitic capacitance and, thus, latency). 
\chyhw{An in-DRAM cache maintains} a copy of a subset of rows from the \chyhw{slow} subarrays in the fast subarrays,
typically caching the \emph{hottest} (i.e., most frequently accessed, \chyhwII{or most recently used}) rows to increase the
probability that a memory request can be served by the fast subarrays \chyhw{(i.e., with low latency)}.

\chyhwII{Unfortunately, existing} in-DRAM cache designs suffer from the inefficient data \chyhw{relocation} operations that copy data between the \chyhw{slow} and fast subarrays.
There are two main reasons for this inefficiency.
First, \emph{the \chyhw{data relocation} granularity is the size of an entire DRAM row}.
In modern DRAM, a row contains a large amount of data (\SI{8}{\kilo\byte} in DDR4~\cite{JEDEC4}).
However, it is difficult for an application to access the entire contents of a row
when the row is opened, as 
\sg{(1)~the application may not have high spatial locality\sgiv{~\cite{ghose.sigmetrics18, ahn.tesseract.isca2015,ahn.pei.isca2015, vijaykumar.isca18}}, and}
(2)~interference among multiple programs running on a multicore processor
limits \chyhw{data reuse from an open row}\chyhwII{~\cite{PartialActivation,SubrowBuffer,Micro-Pages, Rowbuffer-Size, ghose.sigmetrics19, Selective-bitline, ChargeCache,CAL,Par-BS,mem-perf-attack,HalfDRAM, ATLAS,MCP,TCM,BLISS,DASH}.}
As a result, when a row in an in-DRAM cache is opened, only a small subset of the cached row
is typically accessed before the row is evicted from the cache.
% \emph{the DRAM-row sized relocation granularity}, because of the inflexible bulk data relocation support, existing in-DRAM data cache designs have to unnecessarily cache an entire DRAM row at a time. Due to the low row buffer locality especially in the multicore systems~\cite{PartialActivation}, most of the cache hits are actually going to only a subset of a cached DRAM row, the rest of the DRAM row is left untouched before evicted out, leading to limited benefit from in-DRAM cache;
Second, \emph{there is a trade-off in current designs between \chyhw{relocation} latency and design complexity}.
% 2) \emph{unscalable relocation latency}, in existing in-DRAM cache designs, 
The further away a \chyhw{slow} subarray is physically from the fast subarray, the \sg{higher the latency that} is required to perform the data \chyhw{relocation}. 
To reduce the \chyhw{relocation} latency, \emph{many} fast subarrays are employed and \emph{interleaved} among \chyhw{slow} subarrays (as shown in Figure~\ref{fig:in-DRAM-cache}a), which leads to increased area overhead (e.g., each fast subarray requires its own local row buffer and peripheral circuitry~\cite{SALP,DDMA}) and manufacturing complexity.

To avoid these inefficiencies, we propose a new approach to efficient data \chyhw{relocation} support across subarrays within a DRAM bank that uses (mostly) existing structures within a modern DRAM \chyhwII{device}. 
%\sg{As shown in Figure~\ref{fig:organization}, all of the subarrays in a bank share a single \emph{global row buffer}. The global row buffer serves to connect one cache line's worth of data (in an x8 DRAM chip, 64~bits, as a cache line is distributed across eight chips) from an active local row buffer to the I/O drivers.}
\chyhwII{As shown in Figure~\ref{fig:organization}, all of the subarrays in a bank share a single \emph{global row buffer}. The global row buffer in a bank serves to connect one \sgii{column's} worth of data from an active local row buffer in the same bank to the I/O drivers \sgii{in a chip}. \sgii{Across a \emph{rank} of chips (i.e., a group of chips operating in lockstep), the global row buffers of a single bank can hold one cache line (i.e., 64~bytes) of data.}}
We make the {\bf key observation} that the global row buffer in a DRAM bank is interconnected with \emph{all} of the local row buffers \chyhw{of the subarrays in the bank}. 
% Even though only one local row buffer within a bank is active at a time in existing DRAM chips, by 
\sgiii{By} safely relaxing some constraints in the operation of the DRAM chip, 
we can use the global row \sgiii{buffer}
% \sgii{buffers across a rank of chips} 
to facilitate fine-grained \chyhw{relocation \sgiii{across subarrays \sgii{(i.e., copying only a single column of data in a DRAM chip, which translates to copying a cache block in a DRAM rank)}}}.
Using this insight, we design a substrate called
% build an efficient DRAM substrate to facilitate cacheblock level (i.e., 64-byte) data relocation across subarrays at a distance independent latency. We call this substrate 
\sg{\emph{\substrate{}}}. 
\substrate{} operations are performed by enabling two local row buffers to transfer data in an \emph{unaligned} manner between each other (i.e., data from one column in the source local row buffer can be written to a \emph{different} column in the destination local row buffer) via the global row buffer, without any use of the off-chip memory channel. By making novel use of existing structures within DRAM, we implement \substrate{} with only modest changes to the peripheral logic within DRAM (\textless 0.3\% chip area overhead), without any changes to the cell arrays.

Based on \substrate{}, we propose a fine-grained in-DRAM cache (\chyhw{\emph{\cachename{}}}), as shown in Figure~\ref{fig:in-DRAM-cache}b.
\chyhw{\cachename{}} avoids the pitfalls of state-of-the-art in-DRAM cache designs\chyhwII{~\cite{DyAsy,LISA,AsymBank,TieredDRAM}}. 
The \emph{key idea} of \chyhw{\cachename{}} is to cache only a portion of a DRAM row \sg{(i.e., a \emph{row segment})} via \substrate{}, instead of caching an entire DRAM row at a time.
This \emph{row segment granularity} \chyhwI{based} \chyhwII{caching} approach yields three benefits.
First, it increases the performance of in-DRAM caches, because a single \chyhwII{in-DRAM cache} row \sgii{(in the fast subarray)} can now contain \chyhw{small} fragments of \emph{multiple} DRAM rows that are likely to be accessed before the fragment is evicted from the cache.
By significantly reducing the amount of cache space wasted on unaccessed data, both the in-DRAM cache hit rate and row buffer hit rate increase substantially.
Second, it simplifies the in-DRAM cache design.
\substrate{} has a \emph{distance-independent} relocation latency \chyhw{within a DRAM bank}, reducing the number of fast subarrays needed to keep the latency low \sgii{compared to state-of-the-art in-DRAM caches} (e.g., \chyhw{\cachename{}} provides benefits with only two fast subarrays per bank) and eliminating the need to interleave fast subarrays among \chyhw{slow} subarrays.
Third, it allows in-DRAM caching to provide potential benefit for conventional DRAM chips that contain only \chyhw{slow} subarrays (as shown in Figure~\ref{fig:in-DRAM-cache}c).
Even without \chyhw{subarrays} with lower access latencies,
\substrate{} allows us to use a small number of rows in a \chyhw{slow} subarray to contain the most frequently-accessed fragments of select DRAM rows.  This increases the row buffer hit rate significantly, allowing a greater fraction of memory requests to be served with the lower row hit latency (as a row already open in a row buffer has a lower access latency than an unopened row).
% considerable performance gain can be achieved even by simply reserving a few DRAM rows in a \chyhw{slow} subarray to collocate hot sub-DRAM-rows via \substrate{}.
As we discuss in Section~\ref{sec:FIGCache:discussion}, \chyhw{\cachename{}} can help mitigate security attacks such as \chyhw{RowHammer\chyhwI{~\cite{RowHammer,jeremie.isca20,lucian.SP20,pietro.SP20,mutlu.tcad20,Date17}}} and row buffer based side channel attacks~\cite{DRAMA}, in addition to its performance benefits.

Our evaluations show that on a system with \chyhwI{both} fast subarrays and \chyhw{slow} subarrays \chyhwII{(Figure~\ref{fig:in-DRAM-cache}b)}, \chyhw{\cachename{}} improves performance by 16.3\% and reduces DRAM energy by 7.8\%, on average across \sg{20 eight-core} workloads, over a conventional system without in-DRAM caching. \chyhw{\cachename{}} outperforms a state-of-the-art in-DRAM cache \chyhwII{design~\cite{LISA}, with} an average \chyhw{performance} improvement of \chyhw{4.6\% for 8-core workloads}.
We show that even in a system \emph{without} any fast subarrays \chyhwII{(Figure~\ref{fig:in-DRAM-cache}c)}, if we reserve \sg{64 of the DRAM rows} in \chyhw{a slow subarray as} an in-DRAM cache, \chyhw{\cachename{}} provides considerable performance gain (12.5\% on average).
% can be achieved when implementing the in-DRAM cache by simply reserving a few DRAM rows in \chyhw{slow} subarrays, without introducing heterogeneity into DRAM banks. 
We demonstrate that the performance benefits of \chyhw{\cachename{}} are robust across many system and mechanism parameters (e.g., cache capacity, \chyhw{caching granularity}, replacement policy, hot data identification policy). 
% Finally, we discuss the potential of FG-Cache to mitigate security attacks like row hammer~\cite{RowHammer} and row buffer based side channel attacks~\cite{DRAMA}. 
We conclude that \chyhw{\cachename{}} is a robust and efficient mechanism to reduce DRAM access latency.

We make the following contributions in this work: 

\begin{itemize}[noitemsep, nolistsep, topsep=0pt, leftmargin=*]
 
\item We propose \substrate{}, an efficient substrate that enables \chyhwII{fine} \chyhw{granularity} \sgii{(i.e., column granularity)} data \chyhw{relocation} across subarrays in a memory bank, at \chyhwII{a latency that is independent of the distance of subarrays from each other}. \substrate{} uses (mostly) existing structures in a modern DRAM chip, with its modifications requiring \textless 0.3\% chip area overhead.% to \chyhw{solely} the peripheral logic.
% within a modern DRAM chip by using (mostly) existing structures, and avoids the need to use the off-chip memory channel during data relocation.

\item We propose \chyhw{\cachename{}}, an efficient in-DRAM cache based on \substrate{}. 
\chyhw{\cachename{}} caches fragments of a DRAM row at the granularity of a \chyhwII{row segment, which can be as small as a cache block. Doing so}
significantly improves \chyhw{in-DRAM} caching performance over state-of-the-art
in-DRAM caches.
Unlike prior works, \chyhw{\cachename{}} can \sgiii{be implemented in} DRAM chips with \chyhwII{both} \chyhw{heterogeneous (i.e., slow and fast)} subarrays
and \chyhw{homogeneous (i.e., only slow)} subarrays.
% We show that FG-Cache can achieve: 1) higher performance gain due to increased cache hit and row buffer hit rate; 2) simpler in-DRAM cache design with fewer fast subarrays, which are no longer necessary to be interleaved among \chyhw{slow} subarrays; 3) considerable performance gain even in conventional DRAM chips without heterogeneity (i.e., only \chyhw{slow} subarrays) by simply reserving a few DRAM rows as cache.

\item We comprehensively evaluate the performance and energy efficiency of \chyhw{\cachename{}}. We show that it substantially improves both the performance and energy efficiency of  \chyhwII{single-core and multi-core} systems with DDR4 DRAM, and that it outperforms state-of-the-art in-DRAM caches.
% We also demonstrate that the performance gains are robust across many system and mechanism parameters.

\end{itemize}

%% file: 2_background.tex
\section{Background}
\label{sec:bkgd:org}
\chyhw{
We provide background about DRAM organizations and operations to understand how \substrate{} works. For more \chyhwII{information}, we refer \chyhwII{the} readers to \chyhwII{prior} works that cover DRAM in \chyhwII{detail~\cite{TieredDRAM,AL-DRAM,DDMA,chang.sigmetrics17,CAL,ChargeCache,CROW,LISA,SALP,jamie.isca2013,FLY-DRAM,ghose.sigmetrics19,AmbitBook,EDEN,DRANG,Solar-DRAM,jamie.isca2012,buddyram.cal15,Seshadri.arxiv19,haocong.isca20,Rowclone,samira17,samira16,samira14,SoftMC,Ramulator}}}.

{\bf DRAM Organization.} A modern main memory subsystem consists of one or more memory \emph{channels}, where each channel contains a \emph{memory controller} that manages a dedicated subset of DRAM modules. The modules in a single channel share an off-chip bus that is used to issue commands and transfer data between DRAM modules and memory controller, which typically resides in the processor. Each module is made up of multiple DRAM chips, which are grouped into one or more \emph{ranks}. For modern x8 DRAM chips in the same rank, there are typically 8~chips that hold data (with some modules containing an additional chip for error-correcting codes, or \chyhwII{ECC~\cite{minesh.DSN19,yixin16,patel.micro2020,Uksong2014}}). All chips belonging to the same rank operate in \emph{lockstep} (i.e., the same command is \chyhw{issued and} performed by all chips simultaneously), and one row of DRAM cells are distributed across all of the chips within a rank.  The chips in a rank, in combination, provide 64~bytes of data (and 8 bytes of ECC code for modules with the extra chip) for each memory request. As Figure~\ref{fig:organization} shows, a chip is divided into multiple \emph{banks},
% \footnote{Modern DRAM architectures such as DDR4 DRAM employ bank
% groups~\cite{JEDEC4}, where each bank group contains several banks. Bank grouping is used to provide more banks at low cost for a DRAM module.}
which can serve memory requests (i.e., loads or stores) in parallel and independently of each other. Each bank typically consists of 32--64 two-dimensional arrays of DRAM cells called \textit{subarrays}\chyhwI{~\cite{SALP,Rowclone,kevin.HPCA14,Ambit,Seshadri.arxiv19}}.

\begin{figure}[t] \centering
    \vspace{-0mm}
    \includegraphics[width=\linewidth]{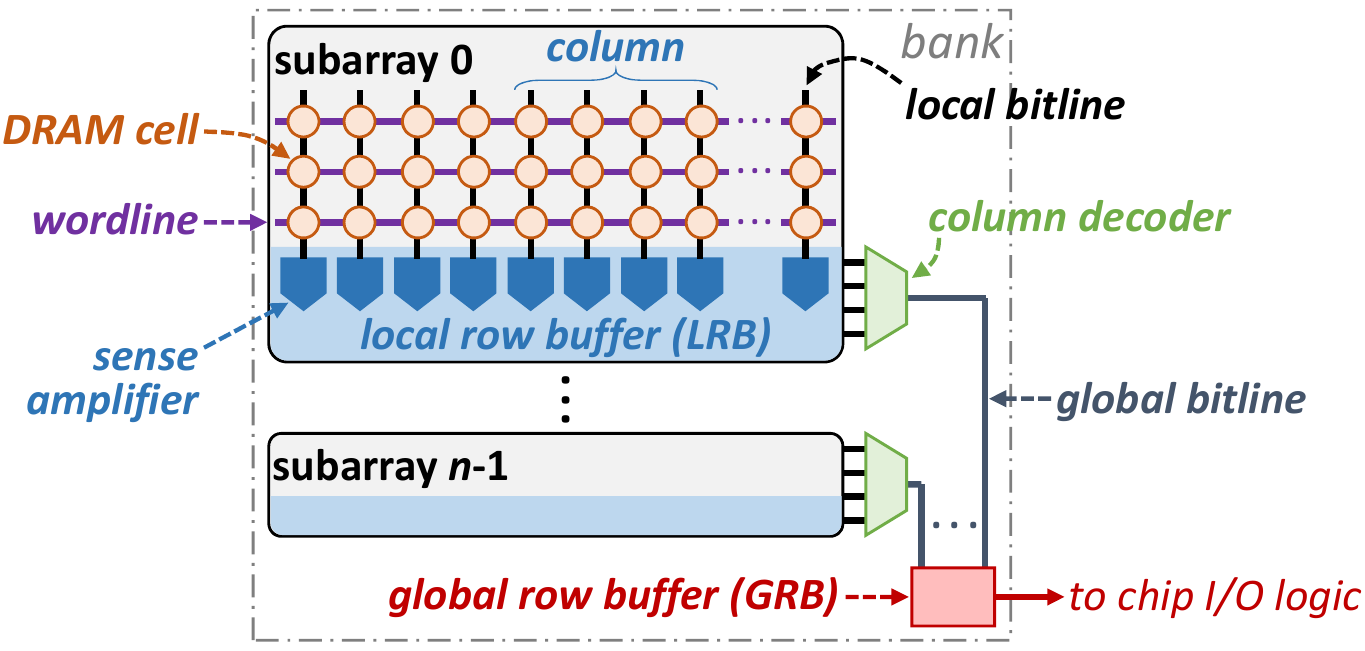}%
    \vspace{-0mm}
    \caption{Detailed DRAM bank and subarray organization.}
    \vspace{-0mm}
    \label{fig:grb}
\end{figure}

In this work, we focus on \chyhw{data movement} operations \emph{across} subarrays within a bank. \chyhw{Figure~\ref{fig:grb} provides more detail about the subarray structure}. Each subarray typically contains 512--2048 rows of DRAM cells, which are connected to a \emph{local row buffer} (LRB). The LRB consists of a set of \emph{sense amplifiers} that are used to open (i.e., activate) one row at a time in the subarray. Each vertical line of cells is connected to one sense amplifier in the LRB via a \emph{local bitline} wire. Cells within a row share a \emph{wordline}. All of the LRBs \chyhwII{in a bank are connected} to a shared \emph{global row buffer} (GRB)~\cite{VLSIMem,VLISMem2,VLISMem3,LISA,SALP}, which is much narrower than the LRB. The GRB is connected to the LRBs using a set of \emph{global bitlines}\chyhw{~\cite{VLSIMem,DDMA}}. The GRB is composed of high-gain sense amplifiers that detect and amplify perturbations caused by a single LRB on the global bitlines~\cite{VLSIMem}. The GRB width is usually correlated with the \chyhw{data output} width of the chip (e.g., in an x8 data/ECC chip, which sends 8~bits of data/ECC for each of the eight data bursts that make up one read, the GRB is 64-bit). 
% Without the GRB, the LRB would either have to take a longer time to drive its values on the I/O, thereby significantly increasing memory latency, or would need to grow significantly in size, leading to scalability issues~\cite{SMLA}. 
Since the GRB is \chyhw{much} narrower than an LRB, a single \emph{column} (i.e., a small number of bits; 64 in an x8 chip) of the LRB is selected using a \emph{column decoder} to connect to the GRB. The column is chosen based on the memory address requested by the DRAM command \chyhwI{that is} being performed.

\label{sec:bkgd:operations}

{\bf DRAM Operations}. The memory controller issues four commands to access and update data within DRAM.  First, the memory controller \emph{activates} the DRAM row containing the data. The ACTIVATE command latches the selected DRAM row into the LRB of the subarray that contains the row. Second, once the activation finishes, the memory controller issues a READ or WRITE command, \chyhw{which operates on a column of data}. On a READ, one column of the LRB is selected using the column decoder and is sent to the GRB via global bitlines. The GRB then drives the data to the chip I/O logic, which sends the data out of the DRAM chip to the memory controller. While a row is activated, the memory controller can issue subsequent READ/WRITE commands to access other columns of data from the LRB if there are other memory requests to the same row. This is called a \emph{row buffer hit}. Finally, the controller \emph{precharges} the LRB \chyhw{and the subarray} by issuing a PRECHARGE command to prepare all of the bitlines for a subsequent ACTIVATE command to a different row. 

The \chyhwII{latencies} of the above commands is governed by timing parameters defined in an industry-wide standard\chyhwI{~\cite{JEDEC,JEDEC4,LPDDR4,TieredDRAM,SALP,DRAMaware-WB}}, which is set largely depending on the length of local bitlines in the subarray. This is because every local bitline has an associated parasitic capacitance whose value is proportional to the length of the bitline. This parasitic capacitance increases the subarray operation latencies during ACTIVATE and PRECHARGE\chyhw{~\cite{TieredDRAM}}.

%% file: 3_existing_in_DRAM_Cache.tex
\section{Existing In-DRAM Cache Designs}
\label{sec:existing}

% Recall from the above description that the local bitline length plays a critical role in the latency of DRAM access. 
DRAM manufacturers often \chyhw{choose a long bitline length} to accommodate a greater number of rows (and, thus, increase DRAM capacity)\chyhw{~\cite{TieredDRAM}}.
To alleviate long DRAM latencies that result from longer bitlines, 
prior works propose in-DRAM caches~\cite{TieredDRAM,AsymBank,DyAsy,LISA}. The key idea of an in-DRAM cache is to introduce heterogeneity into DRAM, where one region has a fast access latency with short local bitline \chyhwI{length}, while the other has bitline length and access latency same as regular \chyhwI{(i.e., slow)} DRAM. To yield the highest performance benefits, the fast region is used as an in-DRAM cache for hot data.  We discuss three main approaches that prior works take in building in-DRAM caches.

\textbf{Heterogeneous Subarray Based Design}. \sgii{Tiered-La\-ten\-cy} (TL) DRAM~\cite{TieredDRAM} divides a subarray into fast (near) and slow (far) segments that have short and long bitlines, respectively, \chyhw{by adding bitline isolation transistors between the two segments}. The fast segment \chyhw{can serve} as an in-DRAM cache. \chyhw{A row can be quickly copied between the two segments via the bitlines, via a back-to-back activation operation resembling RowClone~\cite{Rowclone}.} The main disadvantage \chyhwI{of TL-DRAM}~\cite{TieredDRAM} comes from the intrusive nature of the bitline isolation transistors inside the subarray. Isolation transistors are different from the existing cell access transistors in DRAM, which are specially designed with low leakage~\cite{DRAM-transistor, DyAsy}. When \chyhw{placed} in the middle of \chyhw{a} subarray, isolation transistors \chyhw{require large cost and can affect DRAM yield.} 
For DRAM chips that use the popular open-bitline architecture\chyhwI{~\cite{haocong.isca20,jamie.isca2013,Keeth.book}}, TL-DRAM increases the area overhead significantly (by 3.15\%)~\cite{LISA}. 
As a result, \chyhwI{such} isolation-transistor-based \chyhwI{in-DRAM} cache designs \chyhwI{can potentially} face a \chyhwI{relatively} high barrier to adoption by commercial DRAM vendors.

\textbf{Heterogeneous Bank Based Design Without Data Relocation Support}. CHARM~\cite{AsymBank} introduces heterogeneity \emph{within} each \emph{bank} by designing a few fast subarrays with (1)~short bitlines for faster data sensing, and (2)~close placement to the chip I/O for faster data transfer. Fast subarrays maintain the same cell array structure as traditional DRAM, \chyhw{leading to simple design and with little effect on DRAM yield}. To fully exploit the potential of fast subarrays, CHARM uses an OS-based scheme to statically allocate \chyhw{frequently used} data to fast subarrays based on program profiling information. The main shortcoming of CHARM is that data relocation between fast and \chyhw{slow} subarrays must be done \chyhw{through the memory channel} using the narrow global data bus of DRAM, which incurs high latency and reduces opportunities to use \emph{dynamic} \chyhw{in-DRAM cache} management polices that adapt to dynamic program phase changes (and \chyhw{that} requires more frequent \chyhw{data relocation} than static profiling-based policies). This substantially limits the potential benefits of CHARM, and makes the overall performance gain of in-DRAM cache depend heavily on the effectiveness of the static, \chyhw{profiling based cache} management policy.

\textbf{Heterogeneous Bank Based Design With Bulk Data Relocation Support}. By taking advantage of DRAM structures, DAS-DRAM~\cite{DyAsy} and LISA-VILLA~\cite{LISA} extend the functionality of CHARM~\cite{AsymBank} with in-DRAM bulk data relocation mechanisms.  These mechanisms dynamically relocate data between fast and \chyhw{slow} subarrays without using the narrow global data bus, enabling faster and more efficient relocation. This allows for the efficient implementation of dynamic in-DRAM cache management policies. Specifically, DAS-DRAM enables DRAM row relocation across subarrays in a bank through a row of relocation cells in each subarray. \sgi{The LISA substrate~\cite{LISA} (upon which the LISA-VILLA in-DRAM caching mechanism is built)} further improves the relocation latency with wide inter-subarray links, serving as a direct data relocation path \chyhw{between especially physically-adjacent} subarrays. Unfortunately, the overall performance of state-of-the-art in-DRAM caches is greatly limited by two characteristics of the existing \chyhwII{in-DRAM} data relocation support.

First, the \chyhw{data relocation} granularity is large and fixed (i.e., an entire DRAM row is relocated at a time). Due to the limited row buffer locality exhibited by many programs~\cite{ghose.sigmetrics19}, most of the in-DRAM cache hits are actually to only a small subset of a cached DRAM row, leaving the rest of the DRAM row untouched before the row is evicted from the cache (i.e., most of the row \chyhw{is brought} into the cache without providing any benefit). The interference among \chyhw{concurrently running} programs in a multicore system further hurts the row buffer locality\chyhwII{~\cite{PartialActivation,SubrowBuffer,Micro-Pages, Rowbuffer-Size, ghose.sigmetrics19, Selective-bitline, ChargeCache,CAL,Par-BS,mem-perf-attack,HalfDRAM, ATLAS,MCP,TCM,BLISS,DASH}.}
Thus, caching an entire DRAM row is \chyhw{usually} not necessary and leads to poor utilization of the \chyhwI{in-DRAM cache (i.e., fast subarray)} space. 
% Besides, the row-sized granularity fails to change the accessing pattern to each cached DRAM row, thus cannot fundamentally improve its locality and the corresponding row buffer hit rate. 
% In addition to the poor utilization, by caching an entire DRAM row, existing in-DRAM caches must rely on existing access patterns to the row, and are unable to increase the hit rate for that row compared to a cache-free design.
\sg{Note that \sgi{while} a cached row can take advantage of low latencies in the fast subarray, \sgi{its} row buffer hit rate does not change, as the contents of the \sgi{cached row (and therefore its locality behavior) remain the same as the source row in the slow subarray}.}

Second, \chyhwI{data} relocation latency increases substantially as the physical relocation distance increases. Each relocation requires the relocated row to be written to each intermediate subarray between the source subarray and the destination subarray.
As a result, the further away a \chyhw{slow} subarray is physically from the fast subarray, the higher the latency is for the data relocation \chyhwI{into and out of the in-DRAM cache}. To mitigate this distance-dependent latency, both DAS-DRAM and LISA-VILLA add \emph{multiple} fast subarrays into DRAM banks, physically interleaving the fast subarrays among \chyhw{slow} subarrays to reduce the average distance between a \chyhw{slow} subarray and its closest fast subarray. \chyhwII{Doing so} greatly increases the area overhead (e.g., each new subarray requires additional peripheral circuitry, such as decoders and a local row buffer) and manufacturing complexity. 
% Moreover, the redundancy of fast subarray space is also increased.

As a result, while DAS-DRAM and LISA-VILLA represent the state-of-the-art for in-DRAM caches, their inefficiencies significantly \chyhwI{impact} the benefits and practicality of the mechanisms.

%% file: 4_Relo_Cab_Design.tex
\section{\substrate{} Substrate}
\label{sec:idar}
\label{sec:idar:move}

To solve the inefficiencies of state-of-the-art in-DRAM cache designs, we propose \sg{\emph{Fine-Grained In-DRAM Data Relocation} (\substrate{})}, a new substrate that enables \chyhwII{fine granularity} \chyhw{\sgi{data} relocation across the subarrays in a bank at a distance-independent latency. 
\sgii{\substrate{} can relocate data at the column granularity in a bank \sgv{(i.e., 64~bits in an x8 DRAM chip, which corresponds to 64-byte cache block granularity in a rank)}.}
\sgi{\sgii{\substrate{}} significantly improves} in-DRAM caching in two ways: 
(1)~\sgi{it enables caching} at the granularity of what we call a \emph{row segment} (consisting of one or more contiguous cache blocks), \sgi{and} \chyhwII{thus} a single in-DRAM cache row can now contain row segments from multiple DRAM rows, leading to higher cache utilization and higher row buffer hit rates; and
% (1)~the \emph{row segment} (i.e., a few cache-blocks) caching granularity, where a single cache row in a fast subarray can now contain fragments of multiple DRAM rows, so that both the in-DRAM cache utilization and row buffer hit rate can be substantially increased, providing higher performance gain; and 
(2)~\sgi{it reduces the need for a large number of} fast subarrays per bank (e.g., \sgi{we use only two for our default configuration} in this paper), \sgi{and they} no longer need to be interleaved among normal subarrays, \sgi{leading to low}
% can be employed and it is no longer necessary to interleave them among normal subarrays, substantially reducing the
area overhead and \sgi{low} manufacturing complexity.} 
%\substrate{} makes use of existing structures in a DRAM chip to efficiently move data between subarrays in the same bank.
%Because of the inefficiencies associated with moving an entire row of data (see Section~\ref{sec:existing}), \substrate{} \chyhw{enables moving} data at the granularity of a \emph{row segment} \chyhw{(i.e., sub-DRAM-row)}, which consists of a few (eight in our implementation) contiguous cache blocks. %\todo{expand, and discuss more about why a row segment and not a cache line}

\subsection{\substrate{} Design}
\label{sec:idar:hw}

\substrate{} is built upon the key observation (as we discuss in Section~\ref{sec:bkgd:org}) that 
% each subarray in a DRAM bank has its own private local row buffer (LRB), but all LRBs 
all of the private per-subarray local row buffers (LRBs) in a bank are connected to a single shared global row buffer (GRB). 
By taking advantage of this \chyhw{connectivity}, \substrate{} can perform \chyhwII{column-granularity} data relocation across subarrays at a distance-independent latency, without using the off-chip memory channel.

\textbf{Transferring Data Between Two \chyhw{Local Row Buffers}.}
To relocate data, \substrate{} introduces a new DRAM command, \sgx{\relcommand{}} \chyhwII{(\emph{relocate column})}. \chyhwII{Within} \sgx{a DRAM chip, \relcommand{} copies one column of data from the LRB of one subarray to the LRB of another subarray within the same bank, \sgii{via the GRB}.}
\sgii{Recall from Section~\ref{sec:bkgd:org} that as eight x8 data chips 
work together in lockstep in a rank, the GRB across all chips is 64~bytes,
and, thus, a \relcommand{} command in such a rank-based system copies one cache block (i.e., 64~bytes).}
% For a rank in DDR4 DRAM, this means that a single \relcommand{} command copies one cache block (i.e., 64~bytes) across all chips in the rank.}
% \relcommand{} copies one cache block (i.e., 64~bytes) from one subarray to another subarray within the same bank. 
The \relcommand{} command has two parameters: 
(1)~the source address and (2)~the destination address.
The source address \sgx{in \relcommand{} consists of only the column address (\chyhwII{since} the source is an LRB containing an already-activated row, as we describe below), while} the destination address \chyhw{consists of} the destination \sgx{subarray index} \chyhwII{(since the destination is a not-yet-active LRB),} and the corresponding column address.
Note that our new command can be easily added by using one of the undefined encodings reserved for future use in the DRAM standard~\cite{JEDEC4}, similar to new DRAM commands that have been proposed in previous studies\chyhwI{~\cite{Rowclone,LISA,SALP,Ambit}}.

\begin{figure}[h]
    \centering
    \vspace{-0pt}
    \includegraphics[width=\linewidth]{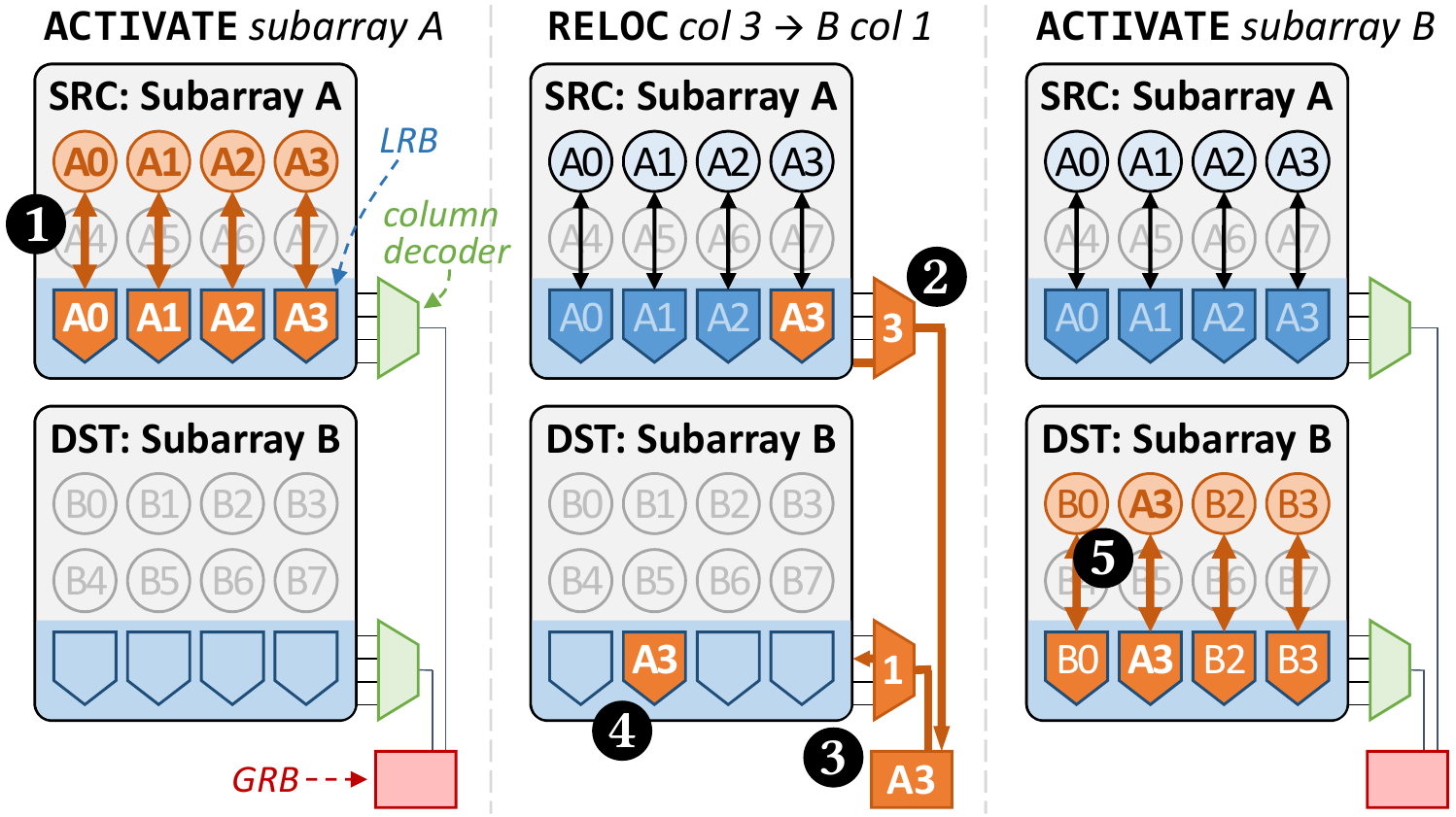}
    \vspace{-12pt}
    \caption{\chyhw{An example of data relocation using \substrate{}}.} %\todo{rename RDMV to \relcommand{}}}
    \label{fig:mig}
    % \vspace{-2pt}
\end{figure}

Figure~\ref{fig:mig} illustrates how \substrate{} relocates \sgx{one column of data} from a source subarray (subarray~A) to a destination subarray (subarray~B) \sgx{in a DRAM chip} using the \relcommand{} command. First, the memory controller issues an ACTIVATE command to one row in subarray A (\circled{1} in Figure~\ref{fig:mig}), which copies data from the selected row to subarray~A's local row buffer (LRB). Second, the memory controller issues a \relcommand{} command. The \relcommand{} command relocates one \sgx{column} of \chyhwII{data (A3 in} Figure~\ref{fig:mig}) from subarray~A's LRB to subarray~B's LRB. To do this, \relcommand{} selects the desired \sgx{column of data} from subarray A's LRB (column~3) using A's column decoder (\circled{2}), which loads the \sgx{column} into the global row buffer (GRB; \circled{3}), and at the same time connects the GRB to subarray B using B's column decoder, which places the \sgx{column of data from subarray~A} in the correct column (column~1) of subarray~B's LRB (\circled{4}). Multiple \relcommand{} commands can be \chyhwII{issued at} this point, copying multiple \sgx{columns} of data from the activated source row to subarray~B's LRB. Third, the memory controller issues an ACTIVATE command to subarray~B (\circled{5}), overwriting only the corresponding column in the activated row with the new \sgx{data (i.e., A3)}. Fourth, the memory controller issues a PRECHARGE command to prepare the \chyhwII{entire} bank for future accesses (not shown in the \sgii{figure).}
% Recall from Section~\ref{sec:bkgd:org} that as eight x8 data chips 
% % and one x8 ECC chip 
% work together in lockstep \chyhw{in a rank}, the GRB across all chips is 64~bytes,
% % (additional 8-byte ECC code is accompanied), so that the 
% and, thus, the granularity of \relcommand{} \sgi{in such a rank-based system} is 64~bytes.
% %\chyhw{in a rank}.

During the \relcommand{} command, \substrate{} relies on the fact that the GRB has a higher drive strength than the LRB~\cite{VLSIMem}. Therefore, when the destination LRB is connected to the GRB \sg{(\circled{3} in Figure~\ref{fig:mig})}, the GRB has enough drive strength to induce charge perturbation to the idle (i.e., precharged) bitlines of the destination subarray, allowing the destination LRB to sense and latch this perturbation even though we are not activating the destination subarray. The GRB will also help to quickly drive the corresponding local sense amplifiers and the bitlines in the destination subarray to a stable state (either $V_{dd}$ or 0). Therefore, when the destination row is activated, the DRAM cells connected to the bitlines in a stable state will be overwritten, while \sgi{all other cells in the row will maintain their original values~\cite{Rowclone,LISA,TieredDRAM} (as seen in \circled{5} \chyhwII{in Figure~\ref{fig:mig})}}. This requires \emph{no} modification \chyhwII{to existing} DRAM. Note that for DRAM modules that contain an additional chip for ECC information, since the data chips and ECC chip operate in lockstep (Section~\ref{sec:bkgd:org}), the corresponding ECC code is transferred together with the data during the relocation \chyhw{process}.
%The memory controller can issue multiple back-to-back \relcommand{} commands to relocate multiple data chunks from the source LRB to the destination LRB, and only activate the destination LRB once all of the \relcommand{} operations are complete for the destination LRB. This can greatly amortize the activation overhead of data relocation.

\textbf{Distance-Independent Latency of \relcommand{}}.
The latency of existing \sgi{data relocation substrates in a DRAM bank}~\cite{LISA,DyAsy} is distance-dependent \chyhw{because \sgi{these substrates} perform} time-consuming sensing of intermediate local row buffers during data relocation, as the data moves from the LRB of one subarray to the next.  As a result, \chyhw{the relocation latency (and energy) is directly dependent on} the number of intermediate local row buffer operations. Unlike existing relocation substrates \sgi{that use local bitlines and isolation transistors for data movement}~\cite{LISA,DyAsy}, the latency of \relcommand{} comes mainly from the sensing of the GRB and the \chyhw{driving of the} destination LRB, \sgii{both via the global bitlines}. While the latency difference between relocation \chyhw{operations} to different subarrays is dependent on the length of global bitlines, a longer global bitline length has a \chyhw{relatively} small impact on the relocation latency, as global bitlines are made of metal with lower capacitance and resistance than local bitlines~\cite{AsymBank}.

Similar to standard READ/WRITE operations in DRAM, whose latencies are set to accommodate worst-case accesses to the furthest subarray, we set the \relcommand{} latency based on the worst case (i.e., \sgi{the latency of} relocating data between two subarrays that are the furthest away from each other \chyhw{when connected} via the global row buffer). We \sgi{use} this worst-case latency \sgi{(plus a safety margin; see Section~\ref{sec:idar:Latency}) as the timing parameter for} \relcommand{}.%, regardless of the distance between source and destination LRBs, since the differences are small.

\textbf{Issuing Multiple Activations Without a Precharge}.
To activate \emph{src} and \emph{dst} rows one after another \emph{without} precharging either row, we must relax the existing constraint that only one row in a bank can be active at a time. Existing memory controllers do \emph{not} allow another ACTIVATE command to be issued to an already-activated bank because the row decoder hierarchy (i.e., the global and local row decoders) \emph{cannot} simultaneously drive two wordlines~\cite{SALP}. While a row is active, the wordline corresponding to the row needs to remain asserted, so that the cells of the row remain connected to the LRB. In existing DRAM chips, the decoder hierarchy latches and drives \emph{only} one-row address, which the memory controller provides along with an ACTIVATE command. To enable \substrate{}, we employ a similar technique to prior work \chyhw{on subarray-level parallelism}~\cite{SALP}\sgi{: we add} a latch \chyhw{to the decoding logic of each subarray} to store an additional row address \chyhw{for holding the source row of \relcommand{}}, and \sgi{extend} the local row decoder of each subarray, such that it can choose between the row address in this latch and the conventional row address bus \chyhw{(to identify the destination row of \relcommand{})}.

\textbf{Enabling Unaligned Data Relocation}.
In a DRAM chip, the column decoder latches and drives \emph{only} one column address per bank, which determines which portion of the LRB is connected to the GRB. 
%For example, in the x8 DRAM chip (i.e., a chip that provides 64~bits for every data burst on the memory channel), the column decoder connects one 64-byte piece of the LRB to the GRB.
Because conventional DRAM activates only one subarray at a time, the column decoder sends a single column address to \emph{all} LRBs in the bank.

To enable \emph{unaligned} data relocation (i.e., relocating data from column~A in the \emph{src} \chyhw{subarray} LRB to column~B in the \emph{dst} subarray \chyhw{LRB}, where $A \neq B$), we need to modify the column decoding logic. When the memory controller sends \emph{two} column addresses simultaneously (one for the source subarray in \relcommand{}, and \chyhw{the} other for the destination subarray), we add a multiplexer to the column decoder of each subarray, to allow the decoder to choose which of the two column addresses it reads (based on whether the subarray contains the \emph{src} or the \emph{dst} LRB). As the existing address bus in DRAM is wide enough to transfer two column addresses at once~\cite{SALP}, we do not need to change the physical DRAM interface.\footnote{\sgi{\relcommand{} uses 21~bits \sg{to express the} column addresses: 7~bits to identify the source column in the open row of the bank, 7~bits to identify the destination subarray index, and 7~bits to identify the destination column.}}%\todo{IMPORTANT: the size and purpose of the sub-DRAM-row has not yet been introduced!}}
% \footnote{\chyhwI{The granularity of \relcommand{} is a cache block, we use always use multiple \relcommand{}s to relocate a row segment at a time,}\sgx{the number of address bits required for \relcommand{} depends on the size of the row segment, as a larger row segment reduces the number of bits needed (but coarsens the relocation granularity). In this work, we empirically choose the size of a row segment to be 16 cache blocks, and so in our implementation, \relcommand{}} uses
% % As described in Section~\ref{sec:FIGCache}, since we always use RDMV to move one row segment (i.e., a few cache blocks) at a time, RDMV requires 
% \sgx{17~bits \sg{to express the} column addresses: 7~bits to identify the destination subarray index, 3~bits \sg{each} to identify the \sg{source and destination row segment indices}, and 4~bits} to identify the cache block within the row segment that is currently being relocated.}%\todo{IMPORTANT: the size and purpose of the sub-DRAM-row has not yet been introduced!}}

\subsection{Latency and Energy Analysis}
\label{sec:idar:Latency}

%In our simulations, we model the entire cell array of a modern DRAM chip using 22 nm DRAM technology parameters, which we obtain by scaling the reference 55 nm technology parameters [89] based on the ITRS roadmap [34, 115].

We perform detailed circuit-level SPICE simulations to find the latency of the \relcommand{} operation.  We analyze a SPICE-level model of the entire cell array of a modern DRAM chip (i.e., row decoder, cell capacitor, access transistor, sense amplifier, bitline capacitor and resistor) with \SI{22}{\nano\meter} DRAM technology parameters, based on an open-source DRAM SPICE model~\cite{CROW}. We use \SI{22}{\nano\meter} PTM low-power transistor models~\cite{PTM22nm1,PTM22nm2} to implement the access transistors and the sense amplifiers. In our SPICE simulations, we run 108~iterations of Monte-Carlo simulations with a $\pm$5\% margin on every parameter of each circuit component, to account for manufacturing process variation and for the worst-case cells in DRAM. Across all iterations, we observe \chyhwII{that \relcommand{}} operates correctly. We report the latency \chyhwII{of \relcommand{} based on} the Monte-Carlo simulation iteration with the highest access latency. 

\begin{figure}[h] \centering
    \vspace{-0pt}
    \includegraphics[width=\linewidth]{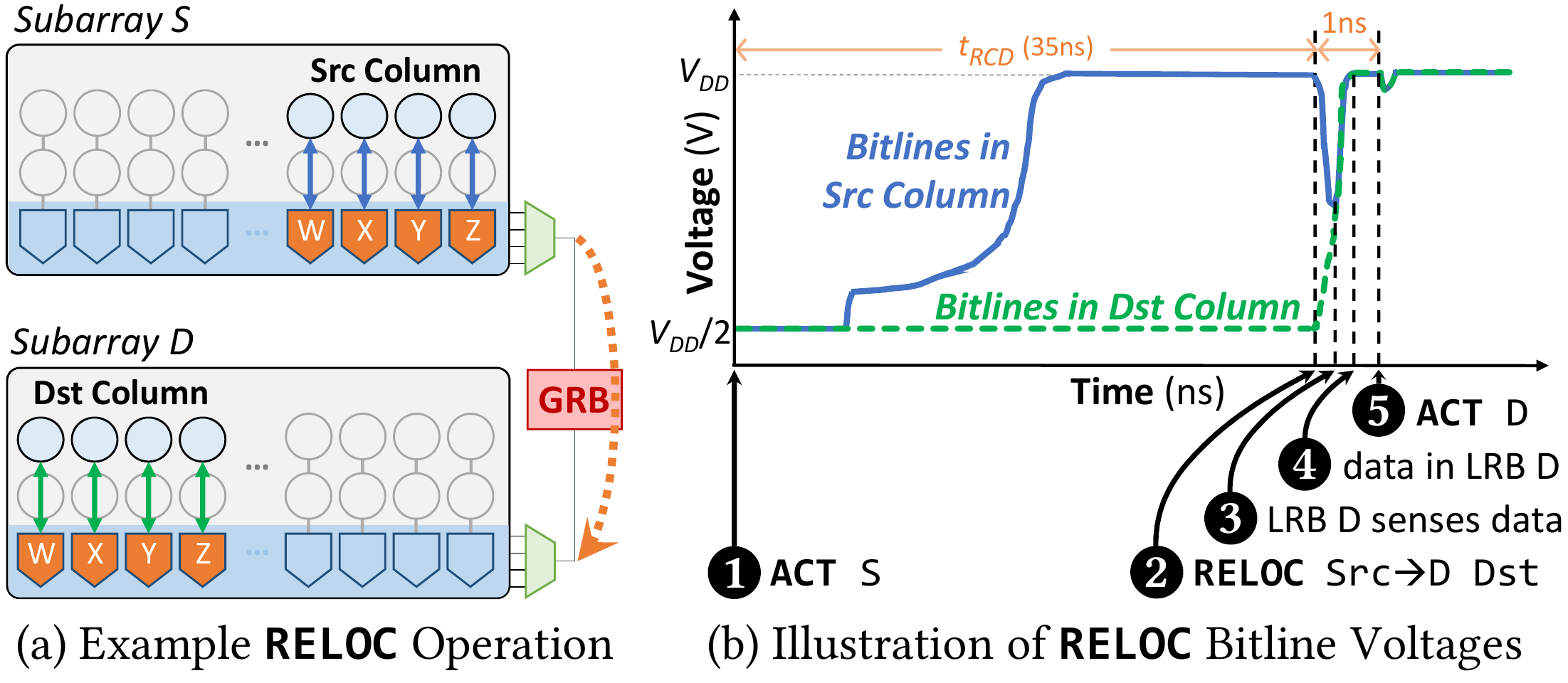}
    \caption{\sgiii{Detailed \relcommand{} \sgiv{operation and timing}}.}% \todo{label each side as a and b}}% \todo{no hyphen in column names; \relcommand{}; voltage graph still ugly}}
    \label{fig:SPICE}
    \vspace{-0mm}
\end{figure}

%We now analyze the process of using one \relcommand{} operation to move data between two subarrays. 
To \sgi{aid} the explanation of \chyhwII{our SPICE simulation of \relcommand{}}, we use an example that performs \relcommand{} from \chyhw{the \emph{Src} column} of LRB~S in subarray~S to \chyhw{the \emph{Dst} column} of LRB~D in subarray~D through the GRB, as shown \sgi{in Figure~\ref{fig:SPICE}a}. 
%\todo{rename the column letters - they are not connected to the subarray ID}.
\sgi{Figure~\ref{fig:SPICE}b} shows the voltage of the bitlines for both \chyhw{the \emph{Src} column (which holds data value 1 \sgi{in each cell})} and \chyhw{the \emph{Dst} column} during the \relcommand{} process over time \chyhw{according to our SPICE simulation}. We explain this \relcommand{} process step by step. 

First, before the \relcommand{} command is issued, an ACTIVATE \sgi{(ACT)} command is sent to subarray~S at time~0 \sgiii{(\circled{1})}. After \SI{35}{\nano\second} (based on the standard-specified $t_{RAS}$ parameter\sgi{~\cite{JEDEC4}}; \sgiii{\circled{2}}), the bitlines are fully restored to $V_{DD}$. Second, the memory controller sends the \relcommand{} command to \chyhw{relocate (copy)} data from LRB~S to LRB~D through the GRB. \relcommand{} turns on the connection between \chyhw{the \emph{Src} column} in LRB~S and \chyhw{the \emph{Dst} column} in LRB~D. Third, after a small amount of time \sgiii{(\circled{3})}, the voltage of the source bitlines in \chyhw{the \emph{Src} column} first drops, as these fully-driven bitlines \chyhw{share charge} with the precharged bitlines \chyhw{in} \chyhw{the \emph{Dst} column} though the GRB. This causes the corresponding sense amplifiers in LRB~D to sense the charge difference and start amplifying the perturbation, during which the GRB helps \sg{amplification} with higher drive strength. In a very short time (less than \SI{1}{\nano\second}), bitlines in \chyhw{the \emph{Dst} column} are fully driven with the value that is originally stored in LRB~S \sgiii{(\circled{4})}. Finally, an ACTIVATE command 
%\todo{is this \circled{4}? if not, show in figure, and explain \circled{4} in text} I modified the figure, we do not need the '4', this is a mistake during figure drawing
is sent to subarray~D \sgiii{(\circled{5})}, overwriting the DRAM cells connected to the bitlines in \chyhw{the \emph{Dst} column},
% , which are in a stable state, will be overwritten, while the others will maintain their
while maintaining the existing values of the other cells in the row~\cite{Rowclone,LISA,TieredDRAM}. %I feel it is not necessary to say the final step, which is not a part of the RELOC command, and can bring additional requirement for the figure.

\sgx{Using SPICE simulations, we find that the latency of \relcommand{}} is \SI{0.57}{\nano\second} (accounting for the worst case of relocating data via the global row buffer between the two subarrays that are the furthest away from each other). 
\sgx{We add a guardband to the \relcommand{} latency, similar to what DRAM manufacturers do to account for process and temperature variation (e.g., the ACTIVATE timing, $t_{RCD}$, has been observed to have extra margins of 13.3\%~\cite{Time-margin} and 17.3\%\chyhw{~\cite{AL-DRAM}).}}
% DRAM manufacturers commonly add a latency guardband in timing parameters to account for process and temperature variation. For instance, the ACTIVATE timing ($t_{RCD}$) has been observed to have extra margins of 13.3\%~\cite{Time-margin} and 17.3\%\chyhw{~\cite{AL-DRAM}} for different types of DRAM devices. 
We add a conservative 43\% guardband for \relcommand{} on top of our SPICE simulation results, resulting in a \SI{1}{\nano\second} latency. 
\sgx{This results in a total latency of \SI{63.5}{\nano\second} to relocate one \chyhwII{column}}
% Overall, the corresponding latency of relocating a 64-byte data element with \substrate{} is \SI{63.5}{\nano\second} 
(i.e., two ACTIVATEs, one \relcommand{}, and one PRECHARGE). %Note that, when relocating multiple consecutive cacheblocks (i.e., a sub-DRAM row), we can issue multiple RDMVs back-to-back, with each cacheblock requiring only an additional 1ns RDMV. 
We estimate the energy consumption of \sgx{a one-cache-block} \sgii{(rank-level)} \substrate{} data relocation operation \sgx{to be \SI{0.03}{\micro\joule},} using the Micron power calculator~\cite{MICRONPower}.
% , and the energy consumption is about \SI{0.03}{\micro\joule} \chyhw{for 64 bytes data relocation}. %Overall, the corresponding latency of relocating a 64-byte data element with our \substrate{} substrate is 63.5ns (i.e., two \texttt{ACTIVATE}s, one \relcommand{}, and one \texttt{PRECHARGE}). 

%% file: 5_FG_Cache_Design.tex
\section{Fine-Grained In-DRAM Cache Design}
\label{sec:FIGCache}
\label{sec:cache}

\substrate{} can improve the efficiency of in-DRAM caches~\cite{LISA,DyAsy} by enabling
(1)~the ability to relocate data into and out of the cache at the \chyhw{fine} granularity of a \emph{row segment} instead of an entire row, \sgi{resulting in higher performance;} and
(2)~designs that %are easier to manufacture and 
avoid the need for a large number of fast \chyhw{(yet low-capacity)} subarrays interleaved among slow subarrays \chyhwI{and thus easier to manufacture}, resulting in lower area overhead \chyhwI{and lower complexity}.
% 2) distance independent data relocation latency, and thus achieving higher performance and lower area overhead and manufacturing complexity.
% %1) sub-DRAM-row (i.e., a few cache-blocks) caching granularity, where a single cache row in a fast subarray can now contain fragments of multiple DRAM rows, so that both the in-DRAM cache utilization and row buffer hit rate can be substantially increased, providing higher performance gain; 2) distance independent data relocation latency, and thus fewer fast subarrays (e.g., only two in this paper) can be employed and it is no longer necessary to interleave them among normal subarrays, substantially reducing the area overhead and manufacturing complexity. 
We use \substrate{} as the foundation of a new in-DRAM cache called \chyhw{\emph{\cachename{}}} (fine-grained in-DRAM cache). %\todo{What about FIGCache?}
\chyhw{\cachename{}} \sgiii{co-locates} hot row segments from slow subarrays into a small number of rows that serve as a cache.
To manage the cache, \chyhw{\cachename{}} uses a \chyhw{tag store (FTS)} \chyhwI{in the memory controller} to hold metadata about currently-cached segments, and employs a simple policy for identifying which segments should be brought into the cache (Section~\ref{sec:existing}). 
When a row segment needs to be brought into the cache, \chyhw{\cachename{}} uses multiple \relcommand{} commands (one for each cache block in the segment) to copy data from the slow subarray into the cache.  Likewise, a dirty evicted row segment is written back \chyhwI{from the cache} to \chyhw{its location in} the slow subarray using \relcommand{} commands.
\chyhw{Rows} serving as the cache can either be implemented using small fast subarrays, reserved rows within \chyhw{slow subarrays}, \chyhwI{or fast rows within a subarray} (Section~\ref{sec:cache:subarray}).
% Note that when relocating a sub-DRAM-row containing multiple cache-blocks, multiple RDMV commands can be issued one after another as soon as the source row is activated (as described in Section~\ref{sec:idar:hw}). \substrate{} also enables us to build in-DRAM cache in conventional DRAM chips with only normal subarrays by simply reserving a few DRAM rows to collocate hot sub-DRAM-rows, not necessarily to introduce fast subarrays. We call this \substrate{} based in-DRAM cache as Fine-Grained In-DRAM cache (\chyhw{\cachename{}}). \chyhw{\cachename{}} contains two main components: a sub-DRAM-row lookup table in the memory controller, and sub-DRAM-row in-DRAM cache in the DRAM device. We describe them below.

\subsection{\chyhw{\cachename{} Tag Store}}
\label{sec:cache:FTS}

\begin{figure}[h]\centering
    \includegraphics[width=1.0\linewidth]{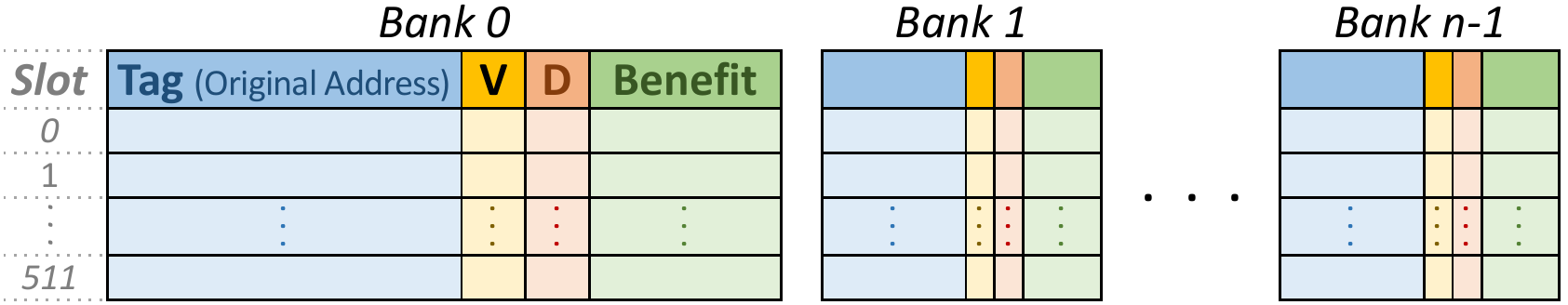}%SLT
    \vspace{-0mm}
    \caption{\sgiii{\cachename{} tag store \sgiv{(FTS)}.}} %\todo{fix sub-row ID and Cnt color}}
    \vspace{-0mm}
    \label{fig:FTS}
\end{figure}

% \textbf{\chyhw{FTS} Structure}. 
\chyhw{A row segment is brought into \cachename{} to lower the latency for subsequent accesses to the segment.} For each memory request, in order to know whether it should be serviced by the cache or by the slow subarrays, the memory controller needs to store information about which row segments are currently \chyhw{cached}. To \chyhwI{this} end, we introduce a \chyhw{\cachename{} tag store (FTS)} in the memory controller. As shown in Figure~\ref{fig:FTS}, 
\sgii{we maintain a fixed \sgiii{portion} of the tag store for each bank. Within each \sgiii{portion}, there is}
% the \chyhw{FTS} is a banked fully-associative structure, containing 
a separate entry for each in-DRAM cache slot \sgii{in the bank} (where each \sgi{fixed-size} slot is the size of one row segment). For each entry, \chyhw{FTS} holds four fields: 
(1)~\sgiii{a tag holding the original address of the row segment}; 
(2)~a valid bit \sgiii{(\emph{V} in the figure)};
% , which indicates whether the caching is valid; 3) the 
(3)~a dirty bit \sgiii{(\emph{D})}; and
% , which indicates whether the data has been modified; and 
(4)~a \emph{benefit} counter \sgiii{(\emph{Benefit})}, which is used for cache replacement. 
% As each \chyhw{FTS} entry corresponds to a unique row segment in the in-DRAM Cache, further indexing into the in-DRAM cache is not needed in the \chyhw{FTS}. We elaborate more on \chyhw{FTS} operations below.  
\sgii{In a bank, \cachename{} acts as a fully-associative cache, and, thus, the entries within each \sgiii{portion} of the FTS are maintained as a fully-associative structure.}

%as previous work~\cite{LISA,TieredDRAM,Micro-Pages} does
\vspace{-0mm}

An \chyhw{FTS} entry is set as valid when a row segment is relocated to the corresponding in-DRAM cache slot.
For every memory request, the memory controller looks up the \chyhw{FTS \sgiii{portion}} \sgiv{associated with the bank of the corresponding request} to determine whether or not \chyhwI{the} request is a hit in the in-DRAM cache. If the request is a \chyhw{\cachename{}} hit (i.e., an \chyhw{FTS} entry matches the row segment ID of the request), its corresponding entry's benefit counter is incremented if the value is not saturated (we empirically set the counter \chyhwI{size} to 5~bits in this work), and the memory controller redirects the request to the in-DRAM cache. 
If the request is a write, the entry's dirty bit is set.
We next discuss how \chyhw{\cachename{}} misses are handled.

% The corresponding dirty bit will be set for a write hit. If there is a miss, \chyhw{FTS} will allocate an entry for the request. This may evict an existing entry, using the replacement policy described below. 
% %The sub-row address is compared against the addresses stored in all entries (e.g., 256 entries in the paper) for that bank. 

\textbf{Choosing \chyhwI{Row} Segments to Insert Into the Cache.}
We rely on a \chyhwI{very} simple policy, \chyhwI{called \emph{insert-any-miss},} to identify when a row segment should be inserted into the cache: every \chyhw{\cachename{}} miss for a memory request  triggers a \chyhwII{row} segment relocation into the cache. This \chyhwII{policy} is designed to achieve the highest utilization of the in-DRAM cache. 
While more sophisticated policies (e.g., adding only those segments whose access frequencies exceed a certain threshold) can be
used to limit the number of segments inserted into \chyhw{\cachename{}}, such policies typically spend additional area and energy on
calculating statistics.
We evaluate the sensitivity of \chyhw{\cachename{}} to various \chyhwII{insertion} policies in Section~\ref{sec:sensitivity}, 
% Other schemes that employ a certain access frequency to a sub-DRAM-row can also be employed to throttle the number of sub-DRAM-row candidates, we have evaluated the sensitivity of Sub-DRAM-row candidate identification policies in Section~\ref{sec:sensitivity}, and shows 
and find that the performance gain of \chyhw{\cachename{}} is robust to different policies,
and that our simple \chyhwI{insert-any-miss} policy performs well.% compared our policy with threshold based one and Note that a large number of works have proposed various identification methodologies (e.g., ~\cite{LISA,Micro-Pages,TieredDRAM}), each of which can be potentially employed, we leave the evaluation of different identification methodologies to the future work.

\textbf{Cache Replacement Policy.}
% 
%To simplify cache management, 
\chyhw{\cachename{}} manages cache replacement at \chyhwI{the} row granularity.
When a new row segment needs to be inserted, and no free segments are available in the cache,
\chyhw{\cachename{}} calculates the cumulative benefit of each \sgiii{in-DRAM cache} row, by summing together the
benefit values of each cached row segment in the row.\sgiv{\footnote{\sgiv{We can use the Dirty-Block Index~\cite{DBI} to simplify the summing operation, as it can help to efficiently maintain per-row benefit scores.}}}
\sgii{The row with the lowest total benefit score is selected for eviction.
\cachename{} maintains a register that holds the ID of the row to be evicted
(6~bits in our configuration), and maintains a single bitvector \sgiii{(8~bits in our configuration)} that tracks
which row segments in the row have not yet been evicted.
When a new \sgiii{in-DRAM cache} row is selected for eviction, the bitvector is set to all ones,
marking all of the row segments in the selected row for eviction.}
% All of the row segments in the row with the lowest total benefit score are marked for eviction.
\sgiii{From the marked row segments, the one} with the lowest individual benefit score is evicted,
making room for the segment being inserted,
\sgii{and its corresponding bit in the bitvector is cleared}.
The other row segments remain marked for eviction \sgii{in the bitvector}, and the next time that a row segment
needs to be inserted, the marked row segment with the lowest score is evicted.
% \todo{we need metadata in the table for this - I do not agree with the explanation in the comments}
% , we do this in the calculation logic, we have cnt in the FTS, we do not need additional tag for this
This process continues for every insertion until no more marked
row segments remain, at which point a new row is selected for eviction.

We choose to perform eviction at a row granularity in order to take advantage of
temporal locality across row segments.
The benefits of \chyhw{\cachename{}} increase when multiple row segments in an open
\chyhw{\cachename{}} row are accessed, as memory accesses to open rows are faster than
memory accesses to closed rows. By evicting all of the segments in a row,
we can pack the row with row segments that are accessed close in time to
each other, increasing the chance (due to locality) that the segments
will again be accessed together, \chyhwI{thus increasing the row buffer hit rate in the in-DRAM cache}.
We compare our row-granularity replacement policy with
commonly-used replacement policies that can be applied at the row segment granularity
in Section~\ref{sec:sensitivity}, and show that our row-granularity replacement
policy achieves higher performance \chyhwI{due to the} higher row buffer hit rate \chyhwI{it enables}.

\subsection{In-DRAM Cache Design}
\label{sec:cache:subarray}

\textbf{Building In-DRAM Cache with Fast Subarrays.}
One way to implement the in-DRAM cache is to add fast subarrays to \chyhwI{a DRAM bank}, \chyhwII{in addition to the regular (i.e., slow) subarrays,} similar to prior works~\cite{AsymBank,DyAsy,LISA}. A fast subarray achieves low access latency by reducing the bitline length\chyhwII{~\cite{TieredDRAM}}. 
Unlike prior works\chyhw{~\cite{AsymBank,DyAsy,LISA}}, whose direct connections between subarrays can incur highly-distance-dependent latencies for \chyhwII{data} relocation (causing designers to interleave many fast subarrays among slow subarrays to \chyhw{bound} the relocation latency),
\substrate{} provides a \emph{distance-independent} \chyhw{relocation latency (Section~\ref{sec:idar:hw})}, as all relocation operations go through \chyhwII{the global} row buffer \chyhwII{and global bitlines that are shared across all subarrays in a bank}.
This allows \chyhw{\cachename{}} to employ only a small number of fast subarrays (we use \chyhwII{only} two per bank in this work), which
reduces both the area \chyhwII{overhead (fewer} subarrays \sgiii{per bank} \chyhwII{lead to fewer peripheral circuitry \sgiii{blocks that are needed for that bank}})  and manufacturing complexity (\sgiv{fewer fast subarrays lead to less design and placement complexity}).
% One important point worth mentioning is that as \substrate{} provides a distance independent latency, it is not necessary to reduce relocation latency between fast and normal subarrays by employing many fast subarrays and interleaving them among normal subarrays. Instead, we can employ fewer fast subarrays (e.g., only two in this paper) than existing in-DRAM cache designs~\cite{DyAsy,LISA}, and arrange these fast subarrays together. This helps to reduce both the area overhead (i.e., fewer fast subarrays) and the manufacturing complexity (i.e., without interleaving fast subarrays).

\textbf{Building In-DRAM Cache with Slow Subarrays.}
As \substrate{} facilitates the \sgiv{co-location} of multiple hot row segments into the same DRAM row, the row buffer hit rate is expected to increase, thus lowering the average DRAM latency. Our row-granularity replacement policy further increases the likelihood of increased row buffer hit rates. As a result, with a low-overhead relocation mechanism, \chyhw{\cachename{}} can improve performance even \emph{without} the aid of reduced-latency subarrays.  This enables us to build \chyhwI{the} in-DRAM cache in conventional homogeneous DRAM chips without introducing heterogeneity into DRAM banks. 

We propose to reserve a \chyhwII{small number of} DRAM rows per bank in a slow subarray, to serve as the in-DRAM cache. Note that DRAM row reservation is a widely-used optimization method in both academia\chyhwII{~\cite{Ambit, ComputeDRAM, Micro-Pages, CROW,buddyram.cal15}} and industry~\cite{Reserved-Rows}. 
One potential issue with using rows in an existing slow subarray is that \substrate{} cannot efficiently relocate data \emph{within the same subarray}.  As a result, to avoid the overheads of relocation, we simply \sgii{do not cache} any row segments from the same subarray that \chyhw{\cachename{}}'s rows reside in.
% One possible concern of the above design is that as \substrate{} cannot efficiently relocation data within the subarray, when the hot sub-DRAM-row and reserved rows are in the same subarray, it cannot be directly relocated via \substrate{}. In this paper, we simply do not relocate the sub-DRAM-row when this happens. 
Given that existing DRAM chips employ a large number of subarrays (i.e., 32 to 64) in each bank~\cite{Ambit,SALP,kevin.HPCA14}, the loss of caching opportunity is negligible. An alternative can be to reserve DRAM rows in two subarrays, and \sgii{relocate} the row segments of one of those subarrays to the reserved rows in the other subarray. However, to simplify the cache management logic, we \sgiii{do not} evaluate such a setup in our work. %Our evaluations show that the reserved DRAM rows based design can achieve considerable performance gains. 

\chyhwI{\textbf{Building In-DRAM Cache with Fast Rows in a Subarray.}}
\chyhwI{Two recent works, CROW~\cite{CROW} and CLR-DRAM~\cite{haocong.isca20},
use the idea of \emph{cell coupling}, where the same bit is written to into two or more cells along the same bitline~\cite{CROW} or wordline~\cite{haocong.isca20} within a subarray.  Cell coupling reduces the access latency when the coupled cells are activated together, as all of the coupled cells now drive their charge simultaneously, increasing the speed at which the data value can be sensed by a sense amplifier. As a result, 
% CLR-DRAM, even adjacent sense amplifiers are coupled, further reducing the access latency. We call 
a row of coupled cells acts as a fast DRAM row, enabling a similar effect as fast subarrays (i.e., low-latency access) without the need for a separate subarray.}
% and a few fast DRAM rows provide a natural substrate as in-DRAM cache in each subarray.}

\chyhwI{Based on the structures proposed in CROW~\cite{CROW} and CLR-DRAM~\cite{haocong.isca20} to write the same bit into multiple cells concurrently, \substrate{} can be extended to relocate data from a conventional slow DRAM row to a fast row. When relocating data from global row buffer to the destination local row buffer, \relcommand{} can utilize the mechanisms proposed in existing works~\cite{CROW,haocong.isca20} so that each bit in the global row buffer can be written into multiple cells (i.e., cells in the fast DRAM row) in the destination subarray. We leave evaluation of such a mechanism to future work.}
%This making these without cell-coupling to the cell-coupling region to fully utilize the cell coupling region.}

\sgii{Tiered-Latency (TL) DRAM~\cite{TieredDRAM} enables fast rows within a subarray by adding isolation transistors
along the bitlines of the subarray. When the isolation transistors are open, only a small number of
rows remain connected to the local row buffer, providing similar performance to a fast subarray.
To build \cachename{} on top of TL-DRAM, we can use \relcommand{} to cache data from the slow rows of
one subarray into the fast rows of a different subarray, as \relcommand{} cannot relocate data
when the source and destination are in the same subarray without incurring additional overheads
(i.e., the  use of a second subarray to serve as an intermediate buffer).
We leave a detailed implementation and evaluation of a TL-DRAM-based \cachename{} to future work.}

\section{\sg{\sgii{Other} Use Cases for \substrate{} and \cachename{}}}
\label{sec:FIGCache:discussion}

%\todo{we need to find a new home for these topics - they come out of nowhere}

\sgii{We believe that \substrate{} and \cachename{} can enable multiple new use cases
(other than \cachename{} in \sgiii{DDRx} DRAM).  We briefly discuss two such cases,
and leave it to future work to design
and evaluate mechanisms that enable these use cases.}

\chyhw{\textbf{\substrate{} with Emerging DRAM Technologies.}} 
Although we evaluate \substrate{} and \chyhw{\cachename{}} for DDR4
DRAM, both solutions can be applied to other DRAM-based
memory technologies with similar bank organizations to
DDR4, such as 3D-stacked High-Bandwidth Memory (HBM)\chyhw{~\cite{HBM,HBM2,SMLA}}
and GDDR5 memory for GPUs\chyhw{~\cite{GDDR5}}.

% Although the current scope of \substrate{} and \chyhw{\cachename{}} only spans within the context of DDR4 standards, these designs are also compatible to other DRAM-based memory technologies, such as 3D-stacked High-Bandwidth Memory (HBM)\chyhw{\cite{HBM,HBM2}} and GDDR5 memory for GPUs\chyhw{\cite{GDDR5}}, which shares similar bank organizations to DDR4.

\chyhw{\textbf{Mitigating DRAM Security \chyhwI{Vulnerabilities} with \cachename{}.}} \chyhw{\cachename{}} can be used to \sgi{reduce the vulnerability of DRAM to} row-buffer-conflict-based attacks. 
\sgii{We \sgiii{briefly examine} two potential vulnerabilities: (1)~RowHammer and (2)~side channel attacks in DRAM.}

RowHammer~\cite{RowHammer,jeremie.isca20,lucian.SP20,pietro.SP20,mutlu.tcad20,Date17} \sgii{is a vulnerability that takes} place when two \chyhw{or more} rows in the same bank are accessed frequently. These frequent accesses cause the two \chyhw{(or more)} rows to be repeatedly open and closed due to row buffer conflicts, hammering (i.e., inducing bit flips in) the data stored in neighboring DRAM rows.
\sgii{\cachename{} helps to reduce the impact of RowHammer because \cachename{} dynamically relocates \sgiii{frequently-accessed} data into a single row.}
Frequently-accessed \sg{row segments} \chyhwI{can be} cached by \chyhw{\cachename{}} \chyhwI{in the same \sgiii{in-DRAM cache} row}, \sg{\sgiv{which eliminates} the need to repeatedly open and close the \sgiii{DRAM} rows that hold each segment}. \sg{\sgiii{\cachename{} reduces} the probability that RowHammer can take place on the in-DRAM cache rows, as \cachename{}'s cache insertion policy keeps row segments accessed around the same time as one another in a single row, significantly reducing the frequency at which multiple \sgiv{in-DRAM} cache rows need to be opened/closed \sgii{(see Section~\ref{sec:eval:overall-performance})}.}

\sgii{A DRAM-row-based side channel attack can be used by a malicious program to locate and monitor the memory accesses of a victim program without the victim's knowledge or permission~\cite{DRAMA}.
In a scenario where the attacker's data \sgiii{is} located in the same bank as data belonging to the} victim, a side channel can be established by monitoring the access time variation caused by row buffer conflicts. 
\sgii{\sgiii{DRAMA~\cite{DRAMA}} demonstrates that this access time variation, coupled with knowledge of where the attacker's data resides in physical memory, can be used to determine when the victim is accessing its data, revealing information such as when a user is performing each keystroke while entering a URL into the address bar of a browser.
The attack works because the attacker can observe information about row hits and misses to specific DRAM rows where its data is co-located with that of the victim.  \cachename{} breaks this ability by caching select portions of DRAM rows, which alters the row hit and miss patterns for the cached data.  Because the attack depends on precise row hit/miss information, \cachename{}'s caching behavior can \sgiii{mitigate} the attack.}
% , intentionally co-locating the data of the attacker and victim in different rows of the same bank will not \sgi{typically} lead to row buffer conflicts as expected, \sgi{eroding} the foundation of \chyhw{the row buffer conflict based side channels}.

\sgiii{We leave the evaluation of both attacks and potential mitigation techniques \sgiv{using \cachename{}} to future work.}

%% file: 6_Methodology.tex
\section{\chyhw{Experimental Methodology}}
\label{Sec:Methodology}

We evaluate \chyhw{\cachename{}} using a modified version of Ramulator\sg{~\cite{Ramulator, ramulator.github}}, a cycle-accurate DRAM simulator, coupled with our in-house processor simulator. We collect user-level application traces using Pin~\cite{Pin}. Table~\ref{tab:system-parameter} shows a summary of our system configuration. We set the \chyhw{default} row segment size as \chyhw{1/8th} of a DRAM row, and study the effect of various row segment sizes on the performance of \chyhw{\cachename{} (Section~\ref{sec:sensitivity})}. For the fast subarray design, we use the open-source SPICE model developed for LISA-VILLA~\cite{LISA}, where slow and fast subarrays have 512 and 32 DRAM rows, respectively, and \chyhw{where timing parameters for activation ($t_{RCD}$), precharge ($t_{RP}$), and restoration ($t_{RAS}$) in fast subarrays can be reduced by  45.5\%, 38.2\%, and 62.9\% respectively}.
% This activate, precharge, and restoration latencies which are reduced from the original timings of normal subarrays by respectively, 45.5\%, 38.2\%, and 62.9\%. 

\begin{table}[h]
    % \vspace{-5pt}%
    \footnotesize
    \begin{center}
     \resizebox{\columnwidth}{!}{
    \begin{tabular}{|c||c|}
        \hline
        % \textbf{Components} & \textbf{Parameters} \\ \hline
        \multirow{2}{*}{\textbf{Processor}} & \chyhwII{8 cores, \SI{3.2}{\giga\hertz}, 3-wide issue, 256-entry \sgiii{inst.} window} \\ 
        & \chyhwII{8 MSHRs/core, L1 4-way \SI{64}{\kilo\byte}, L2 8-way \SI{256}{\kilo\byte}} \\ \hline
        \textbf{Last-Level Cache} & \SI{2}{\mega\byte}/core, \SI{64}{\byte} cache block, 16-way \\ \hline
        \chyhwII{\textbf{Memory Controller}} & \chyhwII{64-entry RD/WR request queues, FR-FCFS scheduling\sg{~\cite{MemSchedule, zuravleff.patent1997}}}\\ \hline
        & DDR4, \SI{800}{\mega\hertz} bus frequency, \\
        & 1 channel for single-core/4 channels for eight-core,\\
        \textbf{DRAM} & 1 rank, 4 bank groups with 4 banks each,\\
        & \sgi{64 subarrays per bank}, \SI{8}{\kilo\byte} row size, \chyhwI{\SI{4}{\giga\byte} capacity per channel}, \\
         & address interleaving: \{row, rank, bankgroup, bank, channel, column\}
        \\ \hline
        \textbf{\chyhw{\substrate{}}} & \chyhwI{\sgii{rank-level} \relcommand{}} granularity: \SI{64}{\byte}, \chyhwI{\relcommand{}} latency: \SI{1}{\nano\second}\\ \hline % \\  & \texttt{\chyhwI{\relcommand{}}} latency: \SI{1}{\nano\second} \\ \hline %energy: \SI{0.03}{\micro\joule}
        & row segment: 1/8th of DRAM row \sg{(16 cache blocks)}, \\  
        \multirow{1}{*}{\textbf{\chyhw{\cachename{}}}} & \chyhwII{fast subarray reduces $t_{RCD}$ / $t_{RP}$ / $t_{RAS}$ by 45.5\% / 38.2\% / 62.9\%\chyhw{~\cite{LISA},}}\\ 
        %& 45.5\% / 38.2\% / 62.9\%\chyhw{~\cite{LISA},}  \\
        & \chyhwII{in-DRAM cache} size: 64~rows per bank \\ \hline
        \textbf{LISA-VILLA} & \chyhwII{in-DRAM cache} size: 512~rows per bank \\ \hline
    \end{tabular}}
    \vspace{-0mm}
    \caption{Simulated system configuration.}
    \label{tab:system-parameter}
    \vspace{-11pt}
    \end{center}
\end{table}

To evaluate energy consumption, we model all major components of our evaluated system based on prior works~\cite{boroumand.asplos18,CAL}, including CPU cores, L1/L2/last-level caches, off-chip interconnects, and DRAM. We use several tools for this, including McPAT 1.0~\cite{mcpat} for the CPU cores, CACTI 6.5~\cite{cacti} for the caches, Orion 3.0~\cite{orion} for the interconnect, and a modified version of DRAMPower~\cite{DRAM-power} for DRAM.

\begin{table}[b]
    % \vspace{-5pt}%
    \footnotesize
    \begin{center}
     \resizebox{\columnwidth}{!}{
    \begin{tabular}{|c|c|}
        \hline
        \textbf{Category} & \textbf{Benchmark Name} \\ \hline
        \multirow{2}{*}{\emph{Memory Intensive}} & zeusmp, leslie3d, mcf, GemsFDTD, libquantum \\
        & bwaves, lbm, com, tigr, mum \\ \hline
        \multirow{2}{*}{\emph{Memory Non-Intensive}} & h264ref, bzip2, gromacs, gcc, bfssandy\\
        & grep, wc-8443, sjeng, tpcc64, tpch2 \\ \hline
    \end{tabular}}
    \vspace{-0mm}
    \caption{Benchmarks used for single-core and multiprogrammed workloads.}
    \label{tab:benchmarks}
    \vspace{-14pt}
    \end{center}
\end{table}

As shown in Table~\ref{tab:benchmarks}, we evaluate twenty single-thread applications from the TPC~\cite{TPC}, MediaBench~\cite{Mediabench}, \sg{Memory Scheduling Championship}~\cite{RT_bench}, Biobench~\cite{Biobench}, and SPEC CPU 2006~\cite{SPEC} benchmark suites. We classify the applications into two categories: memory intensive (greater than 10 last-level cache misses per kilo-instruction, or MPKI) and memory non-intensive (less than 10 MPKI). To evaluate the effect of \chyhw{\cachename{}} on a multicore system, we form 20 eight-core multiprogrammed workloads. We vary the load on the memory system by generating workloads where 25\%, \sgiii{50\%, 75\%,} and 100\% of the applications are memory intensive. 
To \sgiii{demonstrate} the performance improvement of \chyhw{\cachename{}} on multithreaded workloads, we evaluate \emph{canneal} and \emph{fluidanimate} from PARSEC~\cite{Parsec}, and \emph{radix} from SPLASH-2~\cite{Splash2}. 
For both the \chyhw{single-core} applications and eight-core workloads, each core executes at least one billion instructions. We report the instruction-per-cycle (IPC) speedup for \chyhw{single-core} applications, and weighted speedup~\cite{Weighted-speedup} as the system performance metric\chyhwII{~\cite{Eyerman08}} for the eight-core workloads. 
\sgi{For multithreaded workloads, we execute the entire application, and report the improvement in execution time.}

%% file: 7_evaluation.tex
%\begin{figure*}[!ht]
%\minipage{0.32\textwidth}
%  \includegraphics[width=\linewidth]{RELO-Cache-Overall.pdf}
%  \vspace{-5.5mm}
%  \caption{Speedup of RELO-Cache over Baseline}\label{fig:RELO-Cache-Overall}
%\endminipage\hfill
%\minipage{0.32\textwidth}%
%  \includegraphics[width=\linewidth]{RELO-Cache-CacheHit.pdf}
%  \vspace{-5.5mm}
%  \caption{In-DRAM Cache hit rate of LISA-Villa and RELO-Cache}\label{fig:RELO-CacheHit}
%\endminipage\hfill
%\minipage{0.32\textwidth}
%  \vspace{-10pt}
%  \includegraphics[width=\linewidth]{RELO-Cache-RowbufferHit.pdf}
%  \vspace{-5.5mm}
%  \caption{Row buffer hit rate of LISA-Villa and RELO-Cache}\label{fig:RELO-RowBufferHit}
% \vspace{-2.5mm}
%\endminipage
% \vspace{-6.5mm}
%\end{figure*}

\begin{figure*}[b]
  \centering
  \includegraphics[width=0.93\linewidth]{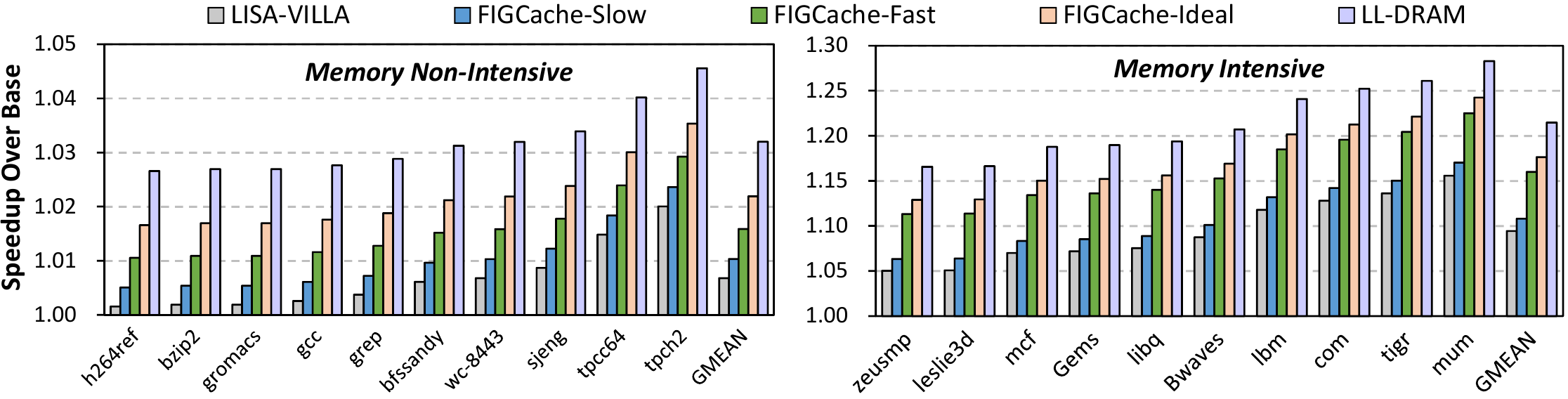}
  \vspace{-5pt}
\caption{\sgiii{Performance of in-DRAM caching mechanisms for single-thread applications, normalized to Base.}}% \todo{fix names and use colors for bars}}
\vspace{-0in}
\label{fig:RELO-Cache-Overall-1core}
\end{figure*}

\begin{figure*}[b]
  \centering
  \includegraphics[width=0.93\linewidth]{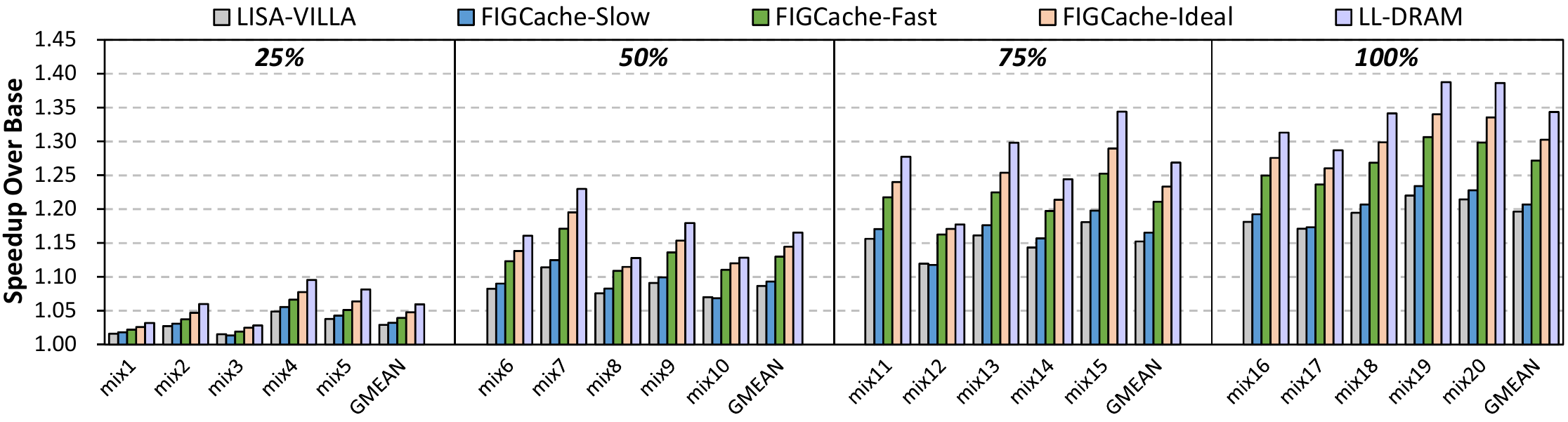}
\vspace{-5pt}
\caption{\sgiii{Performance of in-DRAM caching mechanisms for eight-core multiprogrammed workloads, normalized to Base.}}% \todo{fix names and use colors for bars}}
\vspace{-0in}
\label{fig:RELO-Cache-Overall-8core}
\end{figure*}

\section{Evaluation}
\label{sec:eval}

We evaluate four \sgv{realistic} configurations to understand the benefits of \cachename{}:
\begin{itemize}
    \item \emph{Base}: a \chyhwI{baseline} system with conventional DDR4 DRAM;
    \item \emph{LISA-VILLA}~\cite{LISA}: a state-of-the-art in-DRAM cache;
    \item \emph{\cachename{}-Slow}: our in-DRAM cache with cache rows stored in \chyhwI{64} reserved rows of \chyhwI{one} existing \emph{slow} subarray (i.e., a system with conventional homogeneous DRAM subarrays); %\todo{let's call this \cachename{}-S}.
    \item \emph{\cachename{}-Fast}: our in-DRAM cache with cache rows stored in \chyhwI{\emph{two}} small \emph{fast} subarrays \chyhwII{(with a total of 64 rows)}. %\todo{let's call this \cachename{}-F} and
\end{itemize}
We also \chyhwII{evaluate} \sgi{two idealized configurations} to examine the impact of certain system parameters:
\begin{itemize}
    %\item \emph{\cachename{}\chyhw{--(FZ)}}: an unrealistic version of \cachename{}\chyhw{--F} where the row segment relocation latency is \emph{zero}; and
    %\item \emph{\cachename{}\chyhw{--(FZA)}}: an unrealistic version of \cachename{}\chyhw{--FZ} where \emph{all} subarrays in the DRAM chip are fast.
    \item \emph{\chyhw{\cachename{}-Ideal}}: an unrealistic version of \cachename{}\chyhw{-Fast} where the row segment relocation latency is \emph{zero}; and
    \item \emph{\chyhw{LL-DRAM}}: \sgi{a system} where \emph{all} subarrays in the DRAM \sgi{chips} are fast \sg{(i.e., low latency)}.
\end{itemize}

% In this section, we evaluate both fast subarray and reserved DRAM rows based Fine-Grained Cache designs (i.e., \chyhw{\emph{\cachename{}}}, where in-DRAM cache is built with two fast subarrays per bank with 32 DRAM-rows each; and \chyhw{\emph{\cachename{}}}-r, where in-DRAM cache is built by reserving 64 DRAM-rows in only one subarray per bank). We also show the performance of state-of-the-art in-DRAM cache design: \chyhw{LISA-VILLA}~\cite{LISA}, which caches an entire DRAM row in the fast subaray at a time. In addition, we provide performance comparisons to two idealized versions of \chyhw{\emph{\cachename{}}}: 1) \chyhw{\emph{\cachename{}}}(FR) \emph{assumes}, ideally, that sub-DRAM-rows can be relocated with zero relocation latency; 2) \chyhw{\emph{\cachename{}}}(FD) further assumes that \emph{all} the subarrays have the same latency as fast subarrays.

\subsection{Performance}
\label{sec:eval:overall-performance}

% {\bf Overall Performance.} 
Figures~\ref{fig:RELO-Cache-Overall-1core} and \ref{fig:RELO-Cache-Overall-8core} show the performance improvement over Base for our single-thread applications (using a one-core system) and eight-application multiprogrammed workloads (using an eight-core system), respectively. In both figures, we group the applications and workloads based on memory intensity (see Section~\ref{Sec:Methodology}). We make four observations from the figures.

First, \sg{both \cachename{}-Slow and \chyhwI{\cachename{}-Fast}} \emph{always} \sg{improve} performance over Base. For our single-thread applications, \chyhw{\cachename{}-Fast} provides an average speedup over Base of 1.5\% (up to 2.9\%) for memory non-intensive applications, and 16.1\% (up to 22.5\%) for \sgii{memory intensive} applications. For our multiprogrammed workloads, \chyhw{\cachename{}-Fast} improves the weighted speedup over Base by an average of 3.9\%, 12.9\%, 21.8\%, and 27.1\% for workloads in the 25\%, 50\%, 75\%, and 100\% \sgii{memory intensive} categories, respectively. Across all 20 eight-core workloads, the average performance improvement of \chyhw{\cachename{}-Fast} is 16.3\%. \sg{\cachename{}-Fast achieves speedups for our three multithreaded applications as well (not shown in the figure), with an average improvement of 16.8\% over Base.} 
\sg{Despite not having cache rows with faster access times, \cachename{}-Slow retains a large fraction of the benefits of \cachename{}-Fast, with} \chyhw{an average performance gain of 5.9\% and 12.4\% for single-thread and multiprogrammed workloads, respectively.}

Second, we observe that compared with \chyhw{LISA-VILLA}, which employs 16 fast subarrays and interleaves them among the normal subarrays, \chyhw{\cachename{}-Fast} \chyhwII{\sgiii{provides} 4.7\% higher performance averaged across our 20 eight-core workloads}, despite employing \sgiii{\emph{only two}} fast subarrays. This is because even though \chyhw{\cachename{}-Fast} has \sgiii{much} fewer fast subarrays per bank, \chyhw{\cachename{}-Fast} caches only \chyhwI{1/8th} of a row at a time and \sgiii{co-locates} multiple row segments with high expected temporal locality in a single cache row.
The increased row buffer hit rate \chyhw{in the in-DRAM cache \sgi{(see \chyhwII{analysis} below)} provides most} of \chyhw{\cachename{}-Fast}'s benefits over LISA-VILLA.
These benefits also allow \cachename{}-Slow to outperform LISA-VILLA by \chyhwII{1.9\%} on average \sgii{across all of our} multiprogrammed workloads, \chyhwI{even though} \cachename{}-Slow has \emph{no} fast subarrays at all.
% This allows \chyhw{\emph{\cachename{}}} to use the fast subarray with both higher row buffer hit ratio and capacity efficiency, achieving significantly higher performance. The performance gain of \chyhw{\emph{\cachename{}}}-r also proves this. \chyhw{\emph{\cachename{}}}-r achieves slightly higher performance gain than \chyhw{LISA-VILLA} for most of the workloads. This demonstrates that the idea of in-DRAM cache does not have to work with fast subarrays, and can achieve considerable performance gain by simply reserving a few DRAM rows in each bank. 
We conclude that reducing the granularity of caching and \sgiv{co-locating} multiple row segments into a single cache row is \chyhwI{greatly} effective for improving the performance of an in-DRAM cache.

Third, the benefits of \chyhwI{\cachename{}-Fast} and \chyhw{\cachename{}-Slow} increase as workload memory intensity increases. On average, compared to Base, \chyhw{\cachename{}-Fast} and \chyhw{\cachename{}-Slow} provide 27.1\% and 20.6\% speedup for 100\% \sgii{memory intensive} eight-core workloads, respectively, whereas they achieve more modest speedups of 3.9\% and 3.2\%, respectively, for 25\% \sgii{memory intensive} workloads. \sg{\chyhwI{There} are multiple reasons for the increased benefits for \sgii{memory intensive} \sgi{workloads:} these workloads (1)~}are more likely to generate requests that compete for the same memory bank (i.e., they induce bank conflicts by accessing different rows), which \cachename{} can potentially alleviate by gathering the accessed row segments of each conflicting row into a single cache row; \sg{and
(2)~may in some cases be more sensitive to DRAM latency}.
% the which creates more opportunities to benefit from in-DRAM cache as collocating multiple hot sub-DRAM-rows can potentially alleviate bank conflicts. This can also be proved by the fact that although the 8-core workloads are composed of the same applications as in the 1-core scenario, we observe much higher performance improvements among 8-core workloads. This is because in multi-core scenarios, multiple simultaneously running applications send requests that are likely to interference with each other, exacerbating the bank conflicts.
\sg{The potential correlation between bank conflicts and \cachename{} effectiveness is corroborated} by the fact that our eight-core multiprogrammed workloads achieve much larger performance improvements than our single-core \sgi{applications}.
% , even though the multiprogrammed workloads consist only of bundles of our single-core applications. 
\sg{Individual} applications in multiprogrammed workloads are likely to interfere with each other, thus exacerbating bank conflicts\chyhwII{~\cite{PartialActivation,SubrowBuffer,Micro-Pages, Rowbuffer-Size, ghose.sigmetrics19, Selective-bitline, ChargeCache,CAL,Par-BS,mem-perf-attack,HalfDRAM, ATLAS,MCP,TCM,BLISS,DASH}}\chyhwII{, which \cachename{} can help to alleviate.}

Fourth, \chyhw{\cachename{}-Fast} approaches the ideal performance improvement of \emph{both} \chyhw{\cachename{}-Ideal} and \chyhw{LL-DRAM}, coming within 1.9\% and 4.6\% respectively, on average, for our eight-core system. \sgii{These improvements indicate that the latency \sgiii{of} cache insertion in \cachename{} is low. When a \cachename{} miss occurs, the memory \sgiv{controller} opens the row containing the data that is being requested. While the row is open, \sgiii{the memory controller uses \relcommand{} operations to relocate the row segment} data into the cache.  Since the row is already open, the first ACTIVATE command discussed in Section~\ref{sec:idar:Latency} is \sgiii{not needed, which greatly reduces} the time required for relocation.  \sgiii{The resulting relocation} latency is low enough that \cachename{} can, \sgiii{in many cases,} behave similarly \sgiii{to low-latency} DRAM, without the associated challenges of low-latency DRAM (e.g., small capacity, high cost).}

\sgii{Overall,} we conclude that \chyhw{\cachename{}} significantly reduces DRAM latency and outperforms a state-of-the-art in-DRAM caching \chyhwII{mechanism, \sgiii{while approaching the performance of a low-latency DRAM design} \sgiv{with only fast subarrays}.} % thanks to its efficient implementation. 
% \chyhw{\emph{\cachename{}}}-r also achieves considerable performance gain.

%\todo{what about \cachename{}-r results on their own?  We should add them} added in the first conclusion.

\begin{figure}[h]
	\centering
	\includegraphics[width=1.0\linewidth]{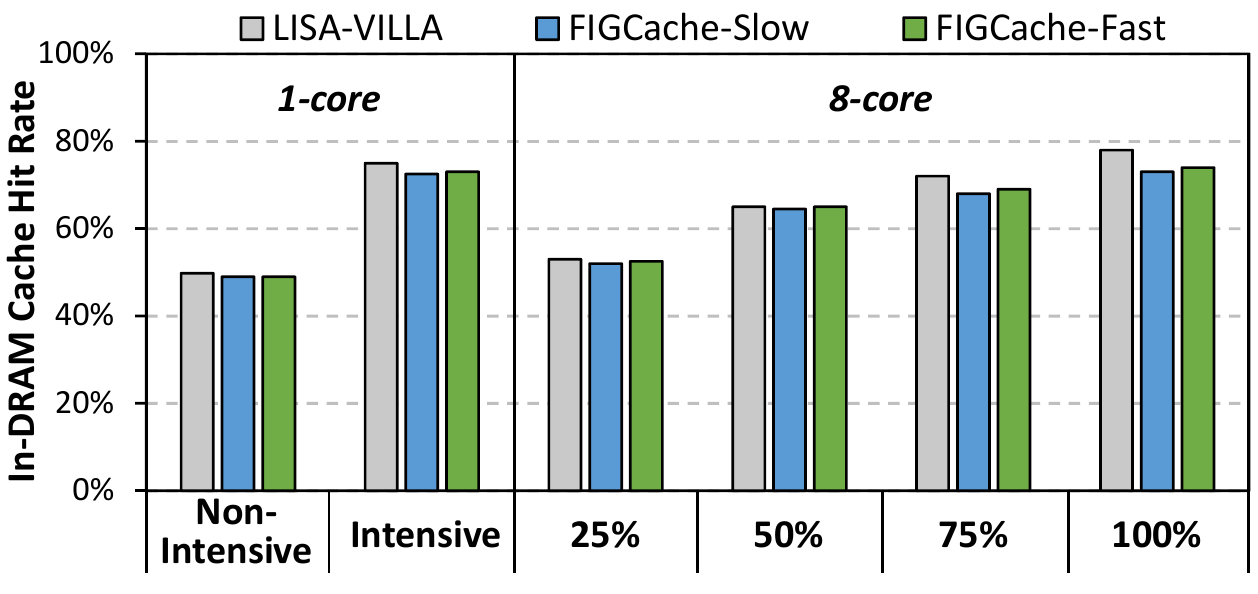}
	\vspace{-12pt}
	\caption{\sgiii{In-DRAM cache hit rate of \chyhw{LISA-VILLA}, \chyhw{\cachename{}-Slow}, and \chyhwI{\cachename{}-Fast}.}}% \todo{fix names and use colors for bars}}
	\vspace{-0mm}
	\label{fig:Cache-Hit-Rate}
\end{figure}

{\bf Cache Hit Rate.}
Figure~\ref{fig:Cache-Hit-Rate} illustrates the in-DRAM cache hit rate 
% (i.e., the fraction of accesses that hit in the fast subarrays (or reserved DRAM rows for \chyhw{\emph{\cachename{}}}-r)) 
of \chyhw{LISA-VILLA}, \chyhw{\cachename{}-Slow}, and \chyhw{\cachename{}-Fast}, averaged across each workload category. We observe that despite having fewer or no fast subarrays, \sg{and having significantly fewer rows reserved for caching,} \chyhw{\cachename{}-Slow} and \chyhw{\cachename{}-Fast} have comparable cache hit rates to \chyhw{LISA-VILLA} across all workloads. This is because due to the limited row buffer locality in many applications, caching an entire DRAM row (as opposed to a row \chyhwII{segment), leads} to \sg{inefficient cache utilization since most of each cached row is not used}. \sg{The finer granularity employed by \cachename{} eliminates much of this inefficient utilization without sacrificing the cache hit rate \chyhwII{with a smaller cache}.} \chyhw{\cachename{}-Slow} \chyhwII{results in} a slightly lower cache hit ratio than \chyhw{\cachename{}-Fast} because, as we discuss in Section~\ref{sec:cache:subarray}, \sg{\cachename{}-Slow} does not cache row segments from the subarray where the reserved rows are allocated.

\begin{figure}[t]
	\centering
	\includegraphics[width=1.0\linewidth]{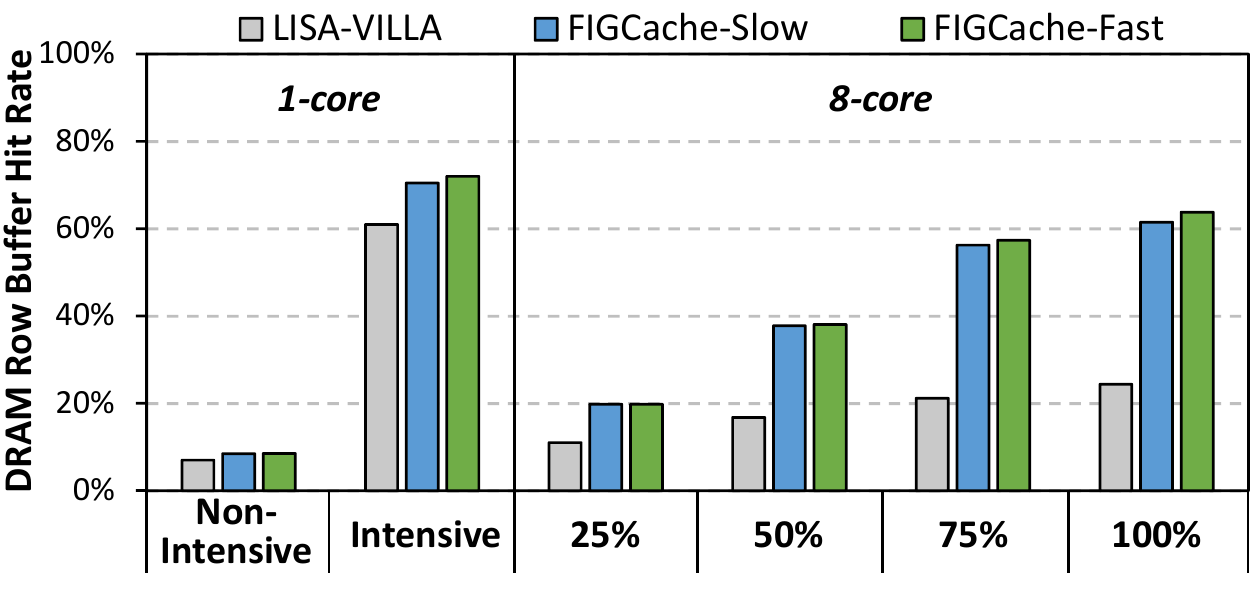}
	\vspace{-12pt}
	\caption{\sgiii{\sgiv{DRAM row} buffer hit rate of \chyhw{LISA-VILLA}, \chyhw{\cachename{}-Slow}, and \chyhwI{\cachename{}-Fast}.}}% \todo{fix names and use colors for bars, hyphen after Non}} %\todo{no hyphen in y-axis label}}
	\vspace{-0mm}
	\label{fig:RB-Hit-Rate}
\end{figure}

{\bf Row Buffer Hit Rate.}
Unlike with the cache hit rate, \chyhw{\cachename{}-Slow} and \chyhw{\cachename{}-Fast} both have significantly-higher \chyhwII{(18\% higher on average) row buffer hit rates for the entire DRAM system than LISA-VILLA,} \sgi{as we observe in Figure~\ref{fig:RB-Hit-Rate}}. 
% As we observe in Figure~\ref{fig:RB-Hit-Rate}, both \chyhw{\cachename{}-Slow} and \chyhw{\cachename{}} have a significantly higher row buffer hit ratio (around 18\% higher on average). 
This is due to two reasons:
(1)~the smaller row segment granularity used by \cachename{}; and
% because \chyhw{\cachename{}-Slow} and \chyhw{\cachename{}} pack many more sub-DRAM-rows into a single DRAM row, achieving a higher row buffer hit ratio.
(2)~our benefit-based cache replacement policy (Section~\ref{sec:cache:FTS}), which increases the row buffer hit \chyhw{rate} by taking into account the temporal locality of multiple row segments during \sgiv{co-location}. 
In contrast, \chyhw{LISA-VILLA} caches an entire DRAM row at a time, and thus the row buffer hit \chyhw{rate} cannot be improved fundamentally beyond the existing row buffer hit rate of the original row.  As a result, LISA-VILLA can benefit \emph{only} from the reduced latencies of a fast subarray.
We conclude that both \chyhw{\cachename{}-Slow} and \chyhw{\cachename{}-Fast} are effective at improving \chyhwII{row buffer hit rate due to their ability to efficiently} \sgiii{co-locate} multiple row segments from different source rows into a single \chyhw{in-DRAM} cache row.

% \textbf{Multithreaded Workloads.}
% To show the performance improvement of FG-Cache on multi-threaded workloads, we evaluate three multi-threaded workloads: Canneal and Fluid from PARSEC~\cite{Parsec}, and RADIX from SPLASH-2~\cite{Splash2}. 
% We evaluate the performance of using \chyhw{\cachename{}} on three multithreaded workloads: Canneal, Fluid, and RADIX. We find that \cachename{} achieves speedups of 15.2\%, 26.3\%, and 9.5\%, respectively for the three applications, over Base. Similar to the single-core applications and multiprogrammed workloads, the major factor that affects the magnitude of the performance improvement is the memory intensity.
% Unlike our multiprogrammed workloads, because of the concurrent access patterns and relatively short time between accesses in multithreaded workloads, the activate, precharge and restoration latencies become more critical, and thus multithreaded applications tend to benefit more from the reduced timing parameters via \chyhw{\cachename{}}. %\todo{we should add a figure and include -r numbers}

\subsection{System Energy Consumption}
\begin{figure}[h]
	\centering
	\includegraphics[width=1.0\linewidth]{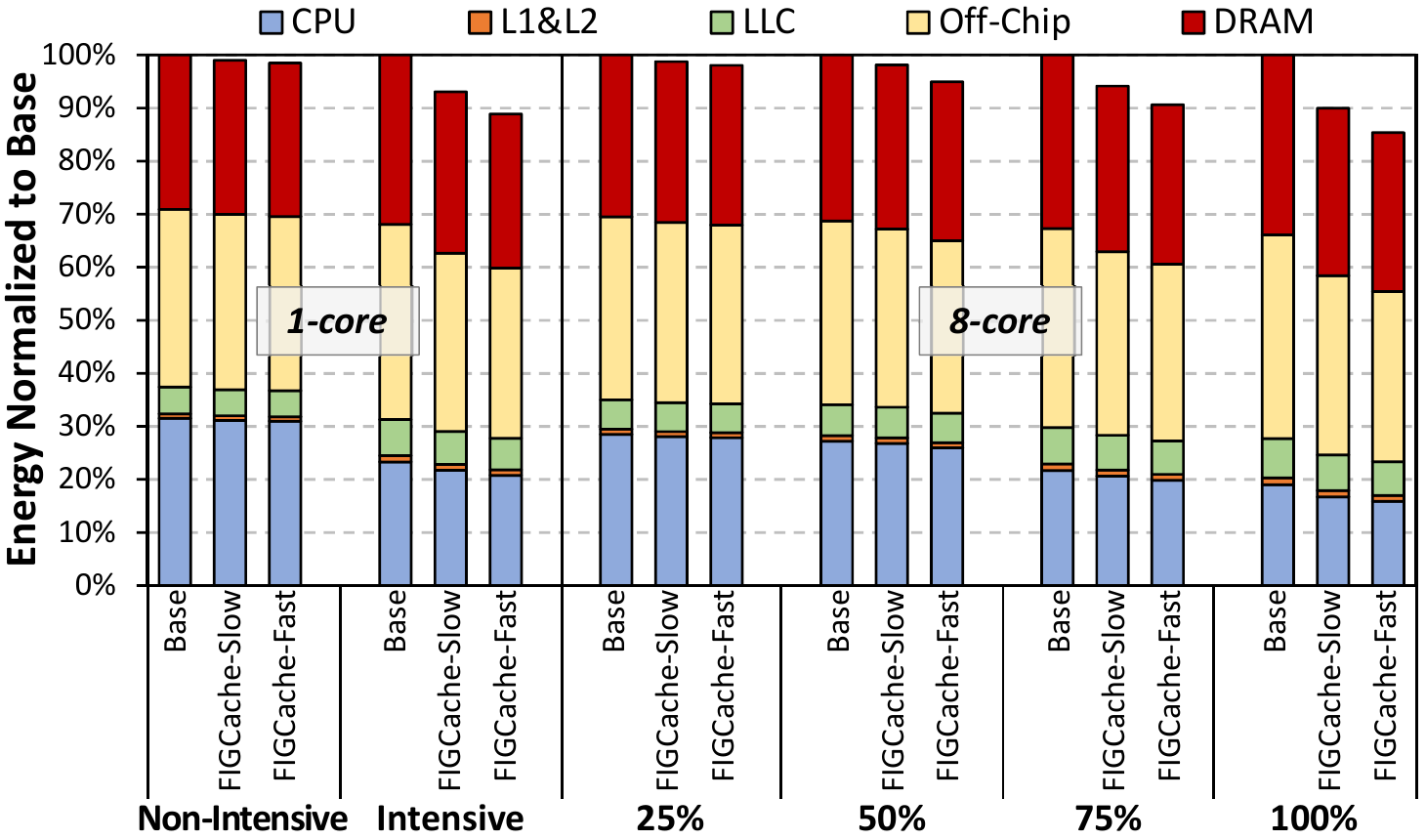}
	\vspace{-3.5mm}
	\caption{\sgiii{Energy \sg{and energy} breakdown of \chyhw{LISA-VILLA}, \chyhw{\cachename{}-Slow}, and \chyhw{\sgiii{\cachename{}-Fast}, \chyhw{normalized to Base}.}}}% \todo{fix names and use colors for bars}}
	\vspace{-0mm}
	\label{fig:energy-breakdown}
\end{figure}

Figure~\ref{fig:energy-breakdown} shows the overall system energy consumption for Base, \chyhw{\cachename{}-Slow}, and \chyhw{\cachename{}-Fast}, averaged across each workload category. We break down the system energy into the energy consumed by the CPU, caches (L1, L2, and LLC), off-chip interconnect (labeled \emph{off-chip} in the figure), and DRAM.

We \chyhwI{draw} two observations from the figure. First, for each workload category, both \chyhw{\cachename{}-Slow} and \chyhw{\cachename{}-Fast} consume less energy than Base. For the \sgii{memory intensive} single-core applications, \chyhw{\cachename{}-Slow} and \chyhw{\cachename{}-Fast} reduce the system energy consumption by an average of 6.9\% and 11.1\%, respectively, compared to Base. 
%The greatest savings are for the 100\% memory-intensive workloads, where the energy is reduced by an average of 14.7\% and 10.1\% over the baseline. 
Second, we observe that the \sgii{energy reduction comes} from two sources: \chyhwII{(1)~improved} row buffer hit rate, which helps to amortize the energy of ACTIVATE and PRECHARGE commands on many memory accesses; and \chyhwII{(2)~reduced} execution time, which saves static energy across each component. For \chyhw{\cachename{}-Fast}, there is a third source of energy reduction, as the faster ACTIVATE and PRECHARGE commands enabled by the fast subarrays further reduce \chyhwII{both dynamic and static} energy. Overall, we conclude that \chyhw{\cachename{}} \chyhw{is} effective \chyhw{at} reducing system energy consumption.

\subsection{Hardware Overhead}
\label{sec:idar:overhead}

\chyhwII{\bf{DRAM Area and Power Overhead}}. 
\substrate{} adds a column address MUX, row address MUX, and a row address latch to each \chyhwI{DRAM} subarray. Our RTL-level evaluation using a \SI{22}{\nano\meter} technology shows that each column MUX occupies an area of \SI{4.7}{\micro\meter\squared} and consumes \SI{2.1}{\micro\watt}, while each row MUX occupies an area of \SI{18.8}{\micro\meter\squared} and consumes \SI{8.4}{\micro\watt}. Each row address latch stores the 40-bit partially predecoded row address, and occupies an area of \SI{35.2}{\micro\meter\squared} with a power consumption of \SI{19.1}{\micro\watt}. 
%The bit selection unit occupies an area of 53.7 um\textsuperscript{2} and consumes 29.3uW. 
For the system configurations described in \chyhwII{Table~\ref{tab:system-parameter}}, the overall area overhead is less than 0.3\% of an entire DRAM \chyhwII{chip.} The overall power consumption is negligible as an activation consumes \SI{51.2}{\milli\watt}~\cite{SALP}. 

\sgiv{\cachename{}-Fast introduces two fast subarrays per bank as an inclusive in-DRAM cache, which is transparent to the operating system. Each fast subarray contains 32~rows (vs.\ 512~rows in each slow subarray). 
Using area estimates from prior works~\cite{DyAsy, LISA}, we calculate that a fast subarray, including cells and sense amplifiers, requires 22.6\% of the area of a slow subarray.
As a result, in our DRAM configuration (see Table~\ref{tab:system-parameter}) where each bank has 64 slow subarrays, the two fast subarrays introduced by \cachename{}-Fast add 0.7\% to the area of the DRAM chip.
In comparison, LISA-VILLA~\cite{LISA} adds 16 fast subarrays to each bank,
which have an area overhead of 5.6\% of the DRAM chip.
\cachename{}-Slow has a lower area overhead than \cachename{}-Fast, as it uses rows
in existing subarrays instead of adding new subarrays, eliminating the area required for
additional sense amplifiers.
As a result, \sgv{\cachename{}-Slow} has an area overhead of only 0.2\% \sgv{in} the DRAM chip.}

% \chyhw{\bf{DRAM Storage Overhead.}} \chyhw{\cachename{}-Fast} introduces two fast subarrays \sgi{per bank} as an inclusive in-DRAM cache, which is transparent to the operating system. Each fast subarray contains 32 rows, and accounts for 1.56\% of the \chyhw{storage overhead} \todo{what does this mean?? see Onur's comments} of \chyhwII{a DRAM chip}. 
% %as the size of the local row buffer (LRB) is much larger than the DRAM rows. 
% Note that \chyhw{\cachename{}-Fast requires} two fast subarrays, while existing in-DRAM designs (e.g., LISA-VILLA~\cite{LISA}) \chyhw{require} more (\sgi{16 fast subarrays per bank for LISA-VILLA, in a bank with 64 slow subarrays}). \chyhw{\cachename{}-Slow} implements an in-DRAM cache using reserved DRAM rows in an existing slow subarray, for which \sgi{we reserve 64 rows per bank (vs.\ 512~rows for LISA-VILLA)}. \chyhwII{The storage overhead \sgiii{of \cachename{}-Slow} is only 0.1\% of the DRAM die.} %accounting for only 0.1\% of the \chyhw{storage} of the DRAM die. 
% %\todo{is this an overhead? don't these already exist?}

\chyhw{\bf{Memory Controller.}}
On the memory controller side, we add the \chyhw{FTS} (Section~\ref{sec:FIGCache}), which incurs modest storage overhead. \sg{We assume \chyhwII{one FTS \sgiii{portion} per bank}, where each \sgiii{portion} has 512 entries.} Each entry of \chyhw{FTS} consists of a row segment address tag, a 5-bit benefit counter, and the dirty and valid bits. The width of the tag is dependent on the number of cached row segments in one bank. For the configuration in Section~\ref{Sec:Methodology}, there are 256K row segments per bank (32K DRAM rows per bank, 8 row segments per DRAM row), which requires a tag size of 19~bits. In total, each entry \chyhw{requires} 26~bits. Therefore, \sg{for each channel in our DRAM configuration (see Table~\ref{tab:system-parameter}), which contains 16~banks with 512 FTS entries per bank, the total storage required for the \chyhw{FTS} is \sg{\SI{26.0}{\kilo\byte}}. Note that compared to \chyhw{LISA-VILLA}~\cite{LISA}, the additional cost of \chyhw{FTS} is only the 3-bit row segment index per entry.}
%(106 kB for the four-channel configuration). 
Using McPAT~\cite{mcpat}, we compute the total area of \chyhwII{all FTS tables} to be \SI{0.496}{\milli\meter\squared} \sgiii{at the \SI{22}{\nano\meter} technology node}, which is only 1.44\% of the area consumed by the \SI{16}{\mega\byte} last-level cache.

We evaluate the access time \sg{and power consumption} of \chyhw{FTS} using CACTI~\cite{cacti}.
\sg{We find that the access time is only \SI{0.11}{\nano\second}, which is small enough that we do not expect it to have a significant} impact on the overall cycle time of the memory controller. \sg{To determine power consumption, we} analyze the \chyhw{FTS} \chyhw{activity} for our applications, accounting for all of the major table operations.  \sgii{Using CACTI~\cite{cacti} and assuming a \SI{22}{\nano\meter} technology \sgiii{node}, we} find that the table consumes \SI{0.187}{\milli\watt} on average. This is only 0.07\% of the average power consumed by the last-level cache. We include this additional power consumption in our system energy evaluations.  
% To keep the design simple, we leave the exploration of \chyhw{FTS} optimizations to future work.

%% file: 8_Sensitivity.tex
\section{Sensitivity \chyhw{Studies}}
\vspace{-0mm}
\label{sec:sensitivity}

In this section, we evaluate our design with various configurations, including different cache capacities, \chyhw{row segment sizes}, cache replacement policies and hot \chyhw{row segment} identification policies. As \chyhw{\cachename{}-Slow} has similar trends with \chyhw{\cachename{}-Fast} for these configurations, we \chyhwII{show results for only} \chyhw{\cachename{}-Fast}. %Unless stated otherwise, we separate the workloads into six categories: 1-core memory non-intensive (Mem-non) and intensive (Memint) workloads, and 8-core workloads with various fraction of memory-intensive applications: 25\%, 50\%, 75\% and 100\%.

\subsection{\sgi{In-DRAM \chyhwI{Cache} Capacity}}

\begin{figure}[h]
	\centering
	\vspace{-0mm}
	\includegraphics[width=1.0\linewidth]{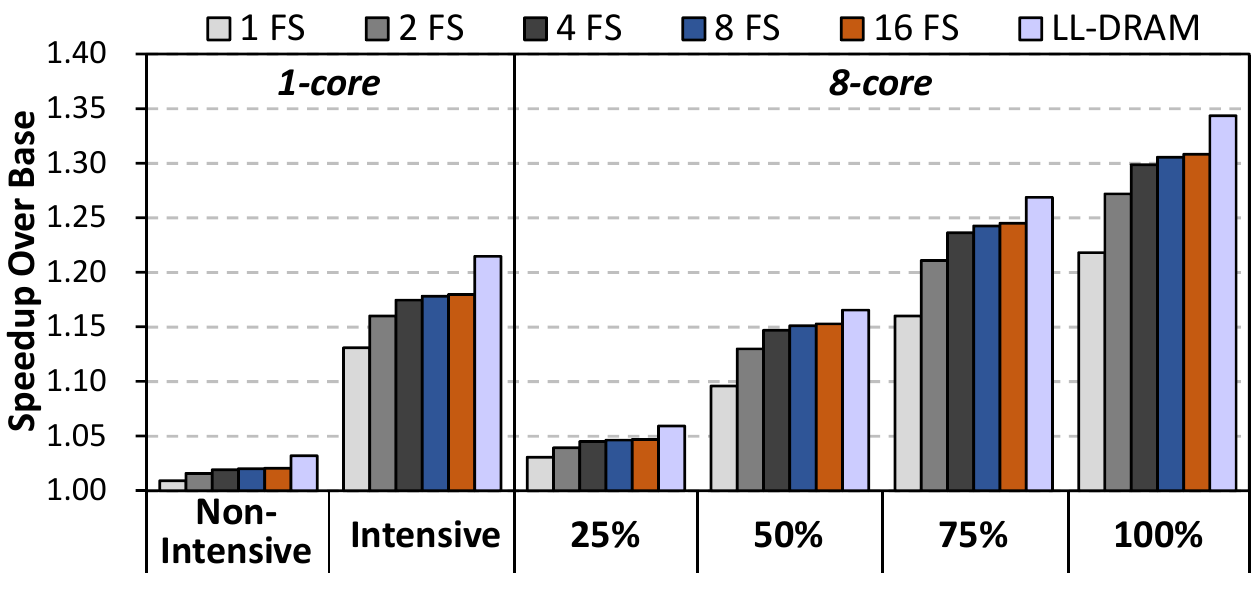}
	%\vspace{-3.5mm}
	\caption{\sgiii{\chyhw{Performance with} different cache capacities.}}% \todo{add LL-DRAM and color, add hyphen after Non}}
	%\vspace{-0.5mm}
	\label{fig:cache-capacity}
\end{figure}

We \chyhw{examine} how the number of fast subarrays in each DRAM bank affects performance. Figure~\ref{fig:cache-capacity} shows the speedup of \chyhw{\cachename{}-Fast} over Base as we vary the number of fast subarrays (\sg{\emph{FS} in the figure}) from 1 to 16. We make two observations from the figure. First, \chyhw{\cachename{}-Fast}’s performance improvement increases with increasing in-DRAM cache capacity. A larger number of fast subarrays reduces the number of evictions, and has the potential to provide more opportunities for \chyhw{\cachename{}-Fast} to reduce access latency for rows that would otherwise be evicted \chyhw{from a smaller in-DRAM cache}. Second, \chyhwII{\sgiii{more} fast subarrays \sgiii{provide diminishing returns} on \cachename{}'s performance improvement, even though \sgiii{they come with} additional storage and complexity overheads.}
%\chyhw{\cachename{}-Fast}’s performance improvement tapers off \chyhw{with a larger number of} fast subarrays. 
For example, increasing the number of fast \chyhwI{subarrays} from 2 to 4 and \chyhw{from 4 to 8 improves performance by less than 2.7\% and 0.8\%, respectively,} for 100\% \sgii{memory intensive} eight-core \chyhwII{workloads.}
%\chyhw{More} fast subarrays require an additional \chyhwI{storage} overhead, but the higher overhead yields diminishing returns.
%, making the additional storage \chyhw{dedicated for the in-DRAM cache}. 
%We conclude that \chyhw{more} fast subarrays provide higher performance, but the performance improvements diminish at larger number. 
\chyhwII{We} implement two fast subarrays per bank to achieve a balance between performance improvement and \chyhw{in-DRAM} storage overhead.

\subsection{\sgi{\chyhw{Row Segment} Size}}

\begin{figure}[t]
	\centering
	\vspace{-0mm}
	\includegraphics[width=\linewidth]{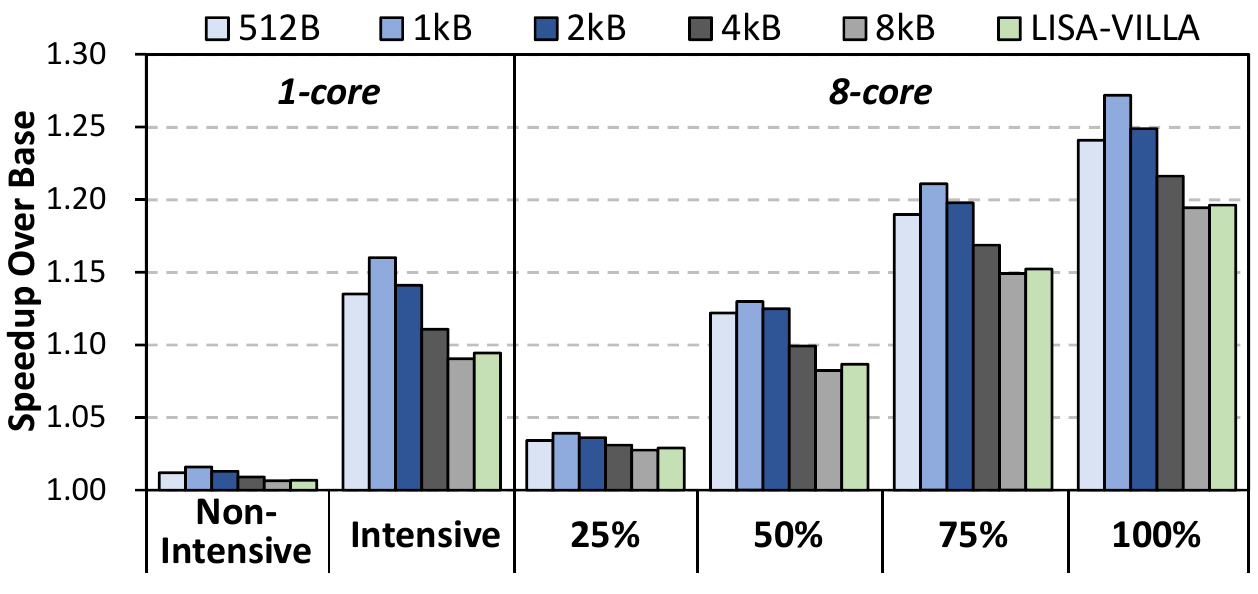}
	\vspace{-10pt}
	\caption{\sgiii{\chyhw{Performance with different row segment} sizes.}}% \todo{add color, express in bytes, sort from smallest to biggest, add LISA}}
	\vspace{-0mm}
	\label{fig:sub-row-size}
\end{figure}

\sg{We vary the size of a row segment to understand its impact on performance.
While a larger row segment size can potentially expose a greater number of opportunities for \chyhwII{exploiting spatial locality within} a DRAM row,
there are \chyhwII{three} downsides:
(1)~many applications do not make use of the contents of an entire row when the row is open,
causing a row segment size that is too large to lead to cache underutilization;
\chyhwII{(2)}~the caching latency increases, as a larger row segment requires more \relcommand{} operations to be issued\chyhwII{; and 
(3)~for a given in-DRAM cache size, a larger row segment size means fewer row segments can be cached.}
Figure~\ref{fig:sub-row-size} shows the performance of \cachename{}-Fast with 
row segment sizes ranging from 8 cache blocks (i.e., \chyhwI{\SI{512}{\byte}, 1/16th} of a DRAM row) to 128 cache blocks (i.e., \chyhwI{\SI{8}{\kilo\byte},} the entire row).
We make two observations from the figure.
First, we find that \cachename{}-Fast performs slightly \emph{worse} than LISA-VILLA~\cite{LISA}
when the row segment size is an entire DRAM row (128 cache blocks).  This is due to the
higher data relocation latency required by \cachename{}, as 128 \relcommand{} operations
are needed, and highlights the benefits of smaller row segment sizes.
Second, we find a peak in performance at a row segment size of 16 cache blocks (i.e., \sgiii{\SI{1}{\kilo\byte}, 1/8th} of a DRAM row), as it
outperforms other row segment sizes across all of our workload categories, and, thus,
we choose this as the row segment size in our implementation.
Note that while we do not evaluate it, \cachename{} can be modified to support 
heterogeneous and/or dynamic row segment sizes (as opposed to the static row segment size that
we currently use).  We leave such a design to future work.}

% \chyhw{We evaluate how the row segment size affects the overall performance by varying its size} from 1-DRAM-row to 1/16 DRAM-row. \chyhw{While a larger row segement} size can potentially capture more of the data locality in a DRAM row, it also introduces longer caching latency (i.e., relocating the \chyhw{row segment} from normal subarray to the fast subarray with more RDMV commands),\footnote{We \chyhw{calculate} the relocation latency for different row segment \chyhw{sizes} according to the number of RDMV commands used during relocation.} and potentially decreases the opportunity to increase row buffer hit rate by combining more \chyhw{row segment}s in the same row. As shown in Figure~\ref{fig:sub-row-size}, for the workloads evaluated in this paper, the sweet-spot \chyhw{for row segment size} is 1/8 DRAM row. When \chyhw{we increase} the \chyhw{row segment} size to an entire DRAM row, \chyhw{\cachename{}'s} performance gain is slightly lower than LISA-VILLA~\cite{LISA}, due to the higher data relocation latency. Note that, \chyhw{\cachename{}-Fast} does not put a strict constraint on \chyhw{row segment} size during data caching: dynamic selection of \chyhw{row segment} size, or having heterogeneous \chyhw{row segment} sizes can be enabled with proper design of the caching mechanisms. We leave this to future work.

\subsection{\sgi{\chyhw{In-DRAM} Cache Replacement Policy}}

\begin{figure}[h]
	\centering
	\vspace{-0mm}
	\includegraphics[width=1.0\linewidth]{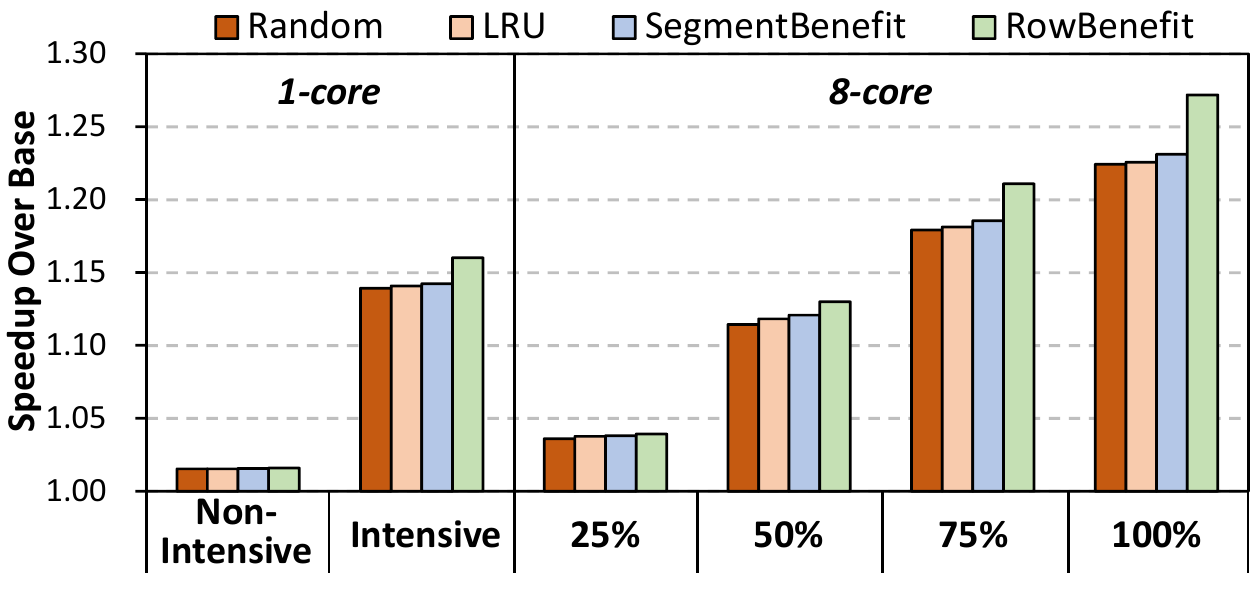}
	\vspace{-10pt}
	\caption{\sgiii{Performance \chyhwI{with} different \sgiv{in-DRAM} cache replacement policies \sgiv{for \cachename{}}.}}% \todo{add color, rename, and keep RowBenefit last}}
	\vspace{-0mm}
	\label{fig:cache-replacement}
\end{figure}

\sg{As we discuss in Section~\ref{sec:cache:FTS}, we implement a new row-granularity benefit-based cache replacement policy for \cachename{}, where the eviction granularity (an entire row) differs from the insertion granularity (a single row segment).
The different eviction and insertion granularities allow us to improve opportunities for \chyhwII{exploiting temporal locality across row segments in an in-DRAM cache row} by packing recently-accessed row segments together into a single cache row.
To understand the benefits of our policy, we evaluate how \cachename{} performs
with three other commonly-used replacement policies.
Figure~\ref{fig:cache-replacement} shows the performance (normalized to Base)
of \cachename{}-Fast using our replacement policy (\emph{RowBenefit} in the figure), 
along with \cachename{}-Fast's performance using:
(1)~\emph{SegmentBenefit}, a traditional benefit-based policy~\cite{TieredDRAM}
where the granularity of eviction is the same as that of insertion (a row segment for \cachename{}),
and only the one row segment with the lowest benefit score anywhere in the in-DRAM cache is evicted;
(2)~\emph{LRU}, a traditional policy that evicts the least-recently-used row segment; and
(3)~\emph{Random}, a policy that evicts a row segment at random from any row in the cache.}
% 
% All of the results presented so far are based on our proposed row-aware-benefit-based cache replacement policy described in Section~\ref{sec:FIGCache}. We \chyhw{evaluate} three other commonly used replacement policies: 1) benefit-based~\cite{TieredDRAM} policy (Benefit), where a benefit counter is employed for each cached \chyhw{row segment}, and incremented whenever the \chyhw{row segment} is accessed. The \chyhw{row segment} with the least benefit value is replaced upon eviction; 2) LRU policy, and 3) Random policy.
% 
% Figure~\ref{fig:cache-replacement} shows the speedup of \chyhw{\cachename{}-Fast} over Baseline with the four replacement policies. We observe that \chyhw{\cachename{}-Fast} provides significant speedups for all three replacement policies. 
% %The speedup of \chyhw{\cachename{}-Fast} is higher for 
% \chyhw{Due to the increased row buffer hit rate, our proposed policy achieves 4.1\% higher performance gain than the best of other policies (i.e. Benefit)} for 100\% memory intensive eight-core workloads. The \emph{Benefit} policy slightly outperforms LRU and random policies, but the difference is negligible. For workloads with lower memory intensity, all policies perform \chyhw{similarly}. Overall, we conclude that \chyhw{\cachename{}-Fast} is effective with different cache replacement policies.

\sg{We make two observations from the figure.
First, \cachename{}-Fast outperforms Base \chyhwII{by more than 12.5\% on average across both \sgii{single-thread and multithreaded} workloads}
with all four cache replacement policies, indicating the benefits of
fine-granularity \chyhwII{in-DRAM} caching regardless of \chyhwII{the exact replacement policy employed}.
Second, our RowBenefit policy either performs the same as, or
outperforms, all three commonly-used policies,
with its benefits increasing as workloads become more
\sgii{memory intensive}.
The RowBenefit policy improves the performance of
\cachename{}-Fast by 4.1\% over the next-best policy (SegmentBenefit)
for 100\% \sgii{memory intensive} eight-core workloads,
due to its increased row buffer hit rate from successfully
improving temporal locality \chyhwII{in in-DRAM cache rows}.
We conclude that 
%\cachename{}-Fast with RowBenefit
\chyhwII{our fine-grained in-DRAM cache with its row-granularity replacement policy}
is effective at capturing temporal locality across
cached row segments.}

\subsection{\sgi{Row Segment Insertion Policy}}

\sg{We use a simple \sgi{\emph{insert-any-miss}} policy to identify which row segments to cache (as we discuss in Section~\ref{sec:cache:FTS}),
where we insert \emph{every} row segment that \chyhwII{misses} in the in-DRAM cache into the cache.
However, it is possible to be more judicious in deciding which row segments should be inserted into the cache.
One example is increasing the threshold of the number of consecutive cache misses \chyhwI{to the row segment}
before \chyhwI{the} segment is inserted.
While a higher threshold can potentially reduce cases where a
row segment is accessed only once across a large time period (in which case
it cannot benefit from caching),
it can also \chyhwII{(1)}~reduce the benefits of caching (by waiting too
long to cache a row segment with high temporal locality), and \chyhwII{(2)~require} additional metadata (as accesses to uncached row segments
now need to be tracked).
To understand the potential benefits of a more judicious insertion policy,
we evaluate different threshold values (where a value of 1 is our
policy of caching a \sgiii{row segment} after \sgiii{a miss to it}), ideally assuming that
the additional storage required for higher thresholds
does not introduce additional latency.}

% We choose a simple hot \chyhw{row segment} identification method in this paper: every miss on the in-DRAM cache (i.e., SLT) triggers the caching of the \chyhw{row segment in the in-DRAM cache}. The policy is designed assuming memory accesses exhibit temporal locality. While adding a threshold to such identification policy can help to throttle the \chyhw{number} of caching candidates, it can also lose the opportunity of in-DRAM caching. To study the effect of this threshold based throttling, we evaluate \chyhw{\cachename{}-Fast} with threshold of 2, 4, and 8, in which a \chyhw{row segment} has to miss in the SLT for 2, 4, and 8 times before being cached.

\begin{figure}[h]
	\centering
	\vspace{-0mm}
	\includegraphics[width=1.0\linewidth]{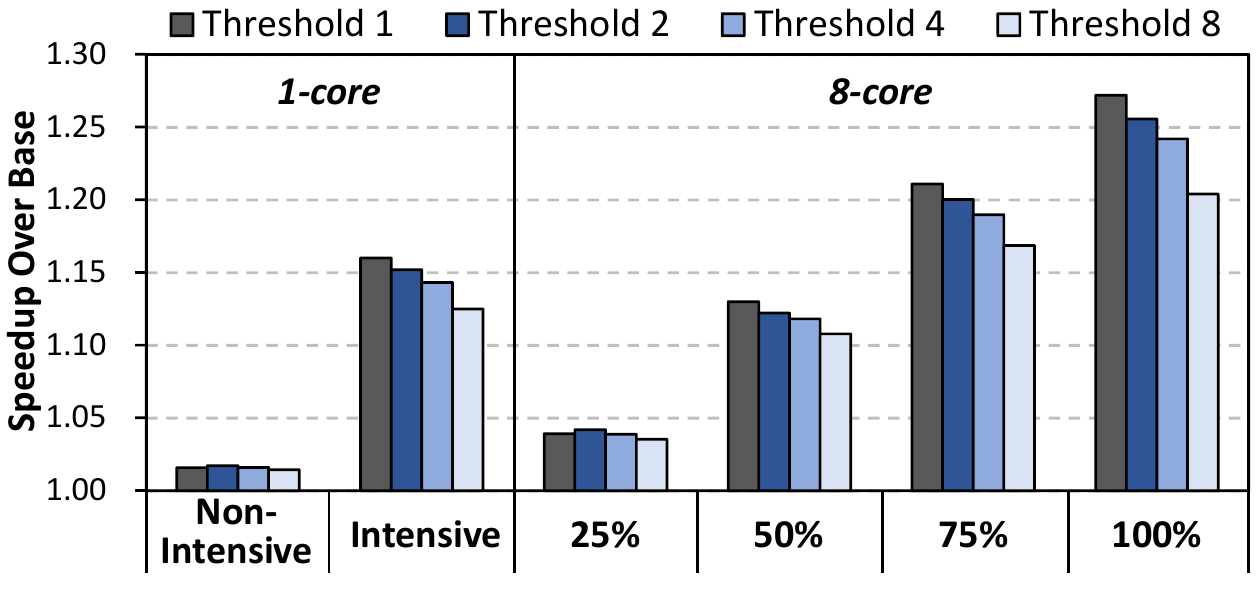}
	\vspace{-10pt}
	\caption{\sgiii{\sg{Performance} \chyhwI{with} different \sg{row segment insertion} thresholds.}}% \todo{add color, typo for Threshold 1}}
	\vspace{-0mm}
	\label{fig:identification-threshold}
\end{figure}

Figure~\ref{fig:identification-threshold} shows \chyhw{\cachename{}-Fast}’s average performance, \sg{normalized to Base, for four threshold values (1, 2, 4, 8). 
We make two observations from the figure.
First, increasing the threshold from 1 to 2 minimally \chyhwI{increases} the performance of memory non-intensive workloads,
though further threshold increases can result in worse performance than a threshold of 1.
Second, for \sgii{memory intensive} workloads, a higher threshold leads to
\emph{worse} performance, by decreasing the number of cache hits \sgii{(latter not shown)}.
% \sgii{Memory intensive} workloads typically experience a greater number of bank conflicts,
% which are mitigated by in-DRAM cache hits (as \cachename{}'s fine granularity
% \chyhwI{collocates different row-segment into the same row in the in-DRAM cache}).  These benefits diminish when there are
% fewer cache hits.
Therefore, we conclude that a threshold of 1 \chyhwI{(i.e., our simple insert-any-miss policy)} is effective for performance.}
% As shown in the figure, increasing the threshold slightly from 1 to 2 can slightly increase the performance gain of memory none-intensive workloads for both single-core and eight-core \chyhw{systems}. However, further increasing the threshold can lead to lower performance gain due to \chyhw{large losses in} potential caching opportunities. While dynamic threshold may help to further increase the overall performance, we set threshold as 1 in this paper to keep the design simple. We conclude that \chyhw{\cachename{}-Fast} effectively improves performance under different thresholds.

%% file: 9_related.tex
\section{Related Work}

To our knowledge, this work is the first to propose \chyhw{an} efficient fine-grained in-DRAM data relocation substrate, which enables \chyhw{a new} \emph{fine-grained} in-DRAM cache design. We already quantitatively demonstrate that \chyhw{\emph{\cachename{}}} outperforms the most closely-related state-of-the-art in-DRAM cache \sg{design, LISA-VILLA~\cite{LISA}}. In this section, we briefly discuss other related works that propose \sg{(1)~other designs for in-DRAM caches, (2)~in-DRAM data relocation support, (3)~\chyhw{designs that improve the} row buffer hit rate; and (4)}~DRAM latency and power reduction mechanisms. 

\textbf{\sg{In-DRAM Caching Mechanisms.}}
\sg{As we discuss in Section~\ref{sec:existing}, there are three main approaches that prior works take
in building in-DRAM caches:
(1)~a heterogeneous subarray based design (Tiered-Latency DRAM~\cite{TieredDRAM}),
(2)~a heterogeneous bank based design without data relocation support (CHARM~\cite{AsymBank}), and
(3)~a heterogeneous bank based design with bulk data relocation support (DAS-DRAM~\cite{DyAsy} and LISA-VILLA~\cite{LISA}).
Like \cachename{}, these works build their in-DRAM caches out of DRAM cells.
Several earlier works\chyhwII{~\cite{kedem.tr1997, hegde.codesisss2003, hidaka.ieeemicro1990, hsu.isca1993, hart.compcon1994,virtual-chan}} on cached DRAM %add rixner's work
integrate SRAM caches into the DRAM modules,
\sgi{usually at very high area overhead~\cite{TieredDRAM, SALP}.}}%\chyhwII{, and \cachename{} can improve the performance of these works due to the increase row buffer hit rate}.}}
% Unlike caches built using DRAM, SRAM-based DRAM caches
% can incur additional latency and/or energy penalties due to their
% need to look up both the SRAM cache and the DRAM cache.
% In contrast, the \cachename{} tag store allows us to perform
% only a single DRAM access, as \cachename{} determines whether
% to send each memory request to a regular row in the slow subarray
% or to a cached row before issuing the request to DRAM.}

\sgi{Similar to traditional caching mechanisms that relocate data into a dedicated cache,
CROW~\cite{CROW}, CLR-DRAM~\cite{haocong.isca20}, and Multiple Clone Row DRAM (MCR-DRAM)~\cite{choi.isca2015}
decrease the access latency for frequently-accessed DRAM rows by coupling multiple cells
together for a single bit of data, thus increasing the amount of charge that is driven to
a sense amplifier when a row is activated.
As we discuss in Section~\ref{sec:cache:subarray}, \cachename{} can be built on top of
the hardware mechanisms that CROW and CLR-DRAM use to manage the fast rows.
While \cachename{} can \chyhwII{also be integrated} with MCR-DRAM, such a design can become more complex,
as MCR-DRAM depends on the OS to manage which pages are assigned to its fast rows~\cite{choi.isca2015}.}

\textbf{In-DRAM Data Relocation Support.}
\sg{The DAS-DRAM~\cite{DyAsy} and LISA~\cite{LISA} substrates provide support for bulk data migration across subarrays, as we discuss in Section~\ref{sec:existing}.}
% Aside from DAS-DRAM~\cite{DyAsy} and LISA-VILLA~\cite{LISA}, which we have discussed previously. RowClone-FPM~\cite{Rowclone} also proposes to perform bulk data relocation support. 
\sg{Another mechanism for bulk data relocation in DRAM is RowClone-FPM~\cite{Rowclone}.}
However, as RowClone-FPM relocates data only within a subarray, it can not be used \sg{to build an in-DRAM cache that caches data from multiple subarrays in a bank.
RowClone-PSM~\cite{Rowclone} is a mechanism that relocates data \chyhwII{at column} granularity across different DRAM \sgii{banks,}
using the shared global data bus inside DRAM (which connects to the memory channel).
Unfortunately, by using the global data bus, RowClone-PSM blocks memory requests to \emph{all} banks during
data relocation, reducing the overall \sgi{bank-level parallelism\sgii{~\cite{Par-BS, lee.micro09}}}.
If RowClone-PSM is used \sgi{to relocate \SI{4}{\kilo\byte} of data} between two subarrays in separate banks, it decreases system performance by 24\% compared to using a conventional \texttt{memcpy} operation~\cite{LISA}.  RowClone-PSM's performance is \chyhwII{even \emph{lower}} for data relocation between subarrays in the \chyhwII{\emph{same}} bank, as this requires two RowClone-PSM operations (one moving data from the source subarray to a second bank that serves as an intermediate buffer, and another moving data from the second bank to the destination subarray in the original bank)~\cite{Rowclone}. \chyhwII{Network-on-Memory (NoM)~\cite{nom.cal20} overcomes this inter-bank limitation of RowClone-PSM with fast \sgiii{and} efficient data relocation across banks within 3D-stacked DRAM, \sgii{via the use of higher connectivity between banks provided by a network in the logic layer}. \substrate{} is \sgiii{orthogonal} to NoM.}}
%, providing more flexible data relocation support together, when implemented in 3D-stacked DRAM.}}

% While RowClone-PSM~\cite{Rowclone} supports cache block \chyhw{granularity} data relocation across memory banks, it has limited efficiency and feasibility \chyhw{because} it blocks all the memory banks from serving other memory accesses during data relocation, this has been proved to degrade system performance by 24\% compared with the traditional memcpy operation when employed for inter subarray data relocation~\cite{LISA}. The situation goes even worse when employed in in-DRAM cache designs, as RowClone-PSM requires a second bank as an intermediate buffer to relocate data from normal to fast subarray, doubling the corresponding latency~\cite{Rowclone}. %\todo{Merge NOM, the bib is:} \chyhw{\cite{nom.cal20}}

\textbf{\chyhwII{Mechanisms} to Improve Row Buffer Hit Rate.}
\sg{Several works mitigate \chyhwII{the negative effects of} low row buffer hit rates by reducing the \sgi{amount of activated data}, either by enabling partial row buffer activation\chyhwII{, designing smaller row buffers, or by in-DRAM data layout \sgii{or transfer} transformations}.}
% A large group of works attempts to enable sub-row-buffer designs to mitigate the low row buffer hit rate problem. 
Examples \sg{of these works} include fine-grained activation~\cite{Fine-Activation}, \sg{Half-DRAM}~\cite{HalfDRAM}, selective bitline activation~\cite{Selective-bitline}, partial row activation~\cite{PartialActivation}, efficient 3D-stacked DRAM designs~\cite{Fine-DRAM,sub-channel-HBM}, \chyhwII{gather-scatter DRAM~\cite{seshadri.micro15}, data reorganization in 3D-stacked DRAM~\cite{DP-accelerate-3D1,DP-accelerate-3D2}}, \sgiv{and row buffer locality aware caching in hybrid memories~\cite{yoon.iccd2012}}.
% These works reduce the side effect of low locality at the DRAM row level by enabling partial row buffer activation or designing smaller row buffers. 
\chyhw{\cachename{}} is orthogonal to these designs, and can be combined with \sgi{them} to \sgi{reduce the amount of unused activated data both in cached rows and in non-cached rows.}
% achieve a better a trade-off between the \sg{row segment size and row} buffer locality. 
At the software level, \sg{prior work proposes to reduce the size of a memory page in the operating system to what \sgi{it calls} micro-pages~\cite{Micro-Pages}, in order to improve spatial locality within a page.  
The reduced page size allows for multiple micro-pages to fit into a single DRAM row, 
and increases the row buffer hit rate by co-locating heavily-used micro-pages into the same row.
While this approach is similar to how \cachename{} collects multiple cached row segments into a
single DRAM row, micro-pages do not have hardware support for relocation, and must
instead use high-latency \texttt{memcpy} operations \sgi{through the memory controller} to relocate data.}
% proposes to reduce the OS page size to micro-pages, and improves the row buffer hit rate by gathering multiple micro-pages into a single DRAM row. In addition to the modification of the OS page size, Micro-pages relocate data using standard \texttt{memcpy} operations, adding considerable overhead that undermines most of its potential benefits from the increased row buffer hit rate.
\sgii{Other techniques to improve the row buffer hit rate include
changing the memory scheduling policy \sgv{(e.g., \cite{Par-BS, TCM, MemSchedule, ipek.isca08, mukundan.hpca12, ghose.isca13, zuravleff.patent1997, ATLAS, ausavarungnirun.isca12, BLISS, yuan.micro09, hur.micro04, nesbit.micro06, DASH, mutlu.micro07, subramanian.iccd14, jog.asplos13, jog.isca13})} to result in more row buffer hits
or introducing \sgiv{new memory allocation policies\sgv{~\cite{regularity-harmful, MCP, das.hpca13, jeong.hpca12, xie.hpca14, yun.rtas14, liu.pact12, jog.asplos13, vijaykumar.isca18, vijaykumar.xmem.isca18}}} to reduce inter-thread interference at the row buffer.
These techniques are orthogonal to \cachename{}.}

\textbf{DRAM Latency and Power Reduction.} 
To reduce DRAM access latency, prior works enable reduced DRAM timing parameters by exploiting the charge level of DRAM cells\chyhwII{~\cite{ChargeCache,NUAT,RestoreTruncation,CAL,DRANG,VRL-DRAM,Reaper,jamie.isca2012}} \sg{or by driving bitlines with charge from multiple cells that contain the same data\sgi{~\cite{CROW, haocong.isca20, choi.isca2015}}}.
%\chyhwI{Two recent works CROW~\cite{CROW} and CLR-DRAM~\cite{haocong.isca20} try to reduce DRAM latency by cell-coupling, that is writing the same bit into two or more adjacent cell along the same bitline~\cite{CROW} or wordline~\cite{haocong.isca20}, this can accelerate the accessing due to that two or more coupled cells are much stronger and thus can be sensed more quickly. \substrate{} is comprehensive to these works, as it can be easily extended to relocate data from a conventional slow region without cell-coupling to the cell-coupling region to fully utilize the cell coupling region}. 
Several other works\sgi{~\cite{Time-margin, AL-DRAM,FLY-DRAM,Solar-DRAM,dlee.sigmetrics17,VRL-DRAM,DL-PUF.hpca2018}} employ optimized timing parameters that take advantage of variation in and across DRAM chips to speed up DRAM accesses. \sg{Aside from} latency reduction, recent studies propose to reduce DRAM row activation and I/O power consumption through efficient row buffer designs (e.g., \sg{multiple sub-row buffers}~\cite{SubrowBuffer}, row buffer caches~\cite{Thread-rowbuffer,RF-rowbuffer,SRAMCache,loh.isca2008}, eager writeback~\cite{eager-RW,EagerWriteback,Virtual-writeQ,DRAMaware-WB}), sub-rank memory~\cite{Sub-rank,rank-subset}, silent writeback elimination~\cite{Write-Value-Locality,Skinflint}, special data encoding schemes~\cite{BusIn-code,Sparse-Codes,Encoding-Similarity, ghose.sigmetrics18}, 
an OS-based scheduler to select different power modes~\cite{Schedule-Energy}, a page-hit-aware low power design~\cite{Power-DRAMDAC08}, and DRAM voltage and/or frequency scaling\sgv{~\cite{MemoryPower, chang.sigmetrics17,mutlu.imw2013,haj-yahya.isca20,deng.asplos11, haj-yahya.hpca20}}. \chyhw{\cachename{}} provides a \chyhw{new} solution for DRAM latency and power reduction, \chyhw{which} can \sg{potentially be} combined with these existing approaches. %\todo{Emerge DRAM Latency PUF, the bib is:} \chyhw{\cite{DL-PUF.hpca2018}} I feel latency PUF is not that related to this work. %All of these approaches are orthogonal and can be applied alongside \chyhw{\emph{\cachename{}}}.

%% file: 10_conclusion.tex
\section{Conclusion}

In this work, we observe that existing in-DRAM cache designs \sg{are inefficient due to (1)~the coarse granularity (i.e., a DRAM row) at which they cache data and (2)~hardware designs that result in high area overhead and manufacturing complexity. 
\sgi{We eliminate} these inefficiencies by introducing}
% suffer from inefficient data relocation mechanisms. We propose 
\substrate{}, a new, low-cost DRAM substrate that enables 
\sg{data relocation (i.e., copying) at the granularity of a \sgi{DRAM column within a chip (cache block within a rank)} with only minor modifications to existing peripheral circuitry in commodity DRAM chips.
Using \substrate{}, we build \cachename{}, a fine-grained in-DRAM cache,}
% cache block \chyhw{granularity} data relocation at a distance independent latency without using the off-chip memory channel. Based on \substrate{}, we build \chyhw{the} fine-grained in-DRAM cache (\chyhw{\emph{\cachename{}}}), 
which greatly improves overall performance \chyhw{and energy consumption}, and 
\sg{has a significantly simpler design than}
% simplifies design of 
existing in-DRAM \chyhw{caches}. 
% We also show that in addition to performance improvements, \chyhw{\emph{\cachename{}}} can be used to prevent security attacks. 
We believe and hope that future works and architectures can exploit the \substrate{} substrate to enable more
\sgi{use cases and \mbox{app\-li\-ca\-tion-/}sys\-tem-le\-vel performance and energy benefits}.
% application \chyhw{and system implementations}.